\documentstyle[aps,pra,tighten,eqsecnum]{revtex}
\begin{document}
\title{Quantum theory of dispersive electromagnetic modes}
\author{P.D. Drummond and M. Hillery}
\address{Department of Physics, The University of Queensland, \\
Queensland 4072, Australia}
\date{\today }
\maketitle

\begin{abstract}
A quantum theory of dispersion for an inhomogeneous solid is obtained,
from a starting point of multipolar coupled atoms interacting with
an electromagnetic field. The dispersion relations obtained are
equivalent to the standard classical Sellmeir equations
obtained from the Drude-Lorentz model. In the homogeneous (plane-wave) case,
we obtain the
detailed quantum mode structure of the coupled
polariton fields, and show that the mode expansion 
in all branches of the dispersion relation is
completely defined by the refractive index and the
group-velocity for the polaritons.
We demonstrate a 
straightforward procedure for 
exactly diagonalizing the Hamiltonian in one, two or three-dimensional
environments,
even in the presence of longitudinal phonon-exciton dispersion,
and an arbitrary number of resonant transitions
with different frequencies.
This is essential, since it is necessary to
include at least one phonon (I.R.) and one exciton (U.V.) mode,
in order to accurately represent dispersion in transparent solid media.
Our method of diagonalization does not require an explicit
solution of the dispersion relation, but relies instead
on the analytic properties of Cauchy contour integrals over
all possible mode frequencies.
When there is longitudinal phonon dispersion, the relevant group-velocity term is
modified so that it only includes the purely electromagnetic
part of the group velocity.
\end{abstract}


\section{Introduction}

Recent developments in quantum optics have led to the realization that
it is essential to include dispersion in the quantum theory of
a linear or nonlinear dielectric, as in a fiber waveguide. 
At the simplest level, it is clear that all dielectric solids have
dispersion and absorption.
There
are an increasing number of experiments that allow direct access to
the quantum nature of propagating radiation fields, ranging
from quantum soliton experiments in optical fibers to QND measurements; as well as
quantum dynamics experiments directed at reducing photon number
noise levels for broad-band communications,
ultra-precise measurements and other device applications \cite{CDRS}-\cite{Friberg}. 
Significantly,
these experiments - as well as more recent theoretical proposals -
have the character of fundamental tests of the quantum theory of
interacting fields \cite{Lieb}-\cite{Lee}, and of quantum measurement theory. 
These are complementary to older
accelerator-based tests,  
taking place in low-energy regimes where a considerable precision
of control is available on the dimensionality of 
the interacting quantum fields. Measurements that can be used
range from the usual particle-counting measurements, to interference
based techniques that allow an unprecedented level of information
about quantum phases. Current technologies even allow measurements
of electromagnetic properties of Bose condensates.

Because the effects of propagation always
involve more than one spatial mode, it is
essential to include dispersion in any physical model of a
linear waveguide in one or more spatial dimensions. 
Treating dispersion can present problems, because it arises
from the fact that the response of the medium to the field
is not instantaneous. The polarization at a given 
time depends not only on the field at that time but on the
values of the field at previous times as well.  This nonlocal
behavior makes standard macroscopic quantization, in which
the medium is represented by its susceptibilities, problematic.
Even without explicit dispersion,  some early treatments
even resulted in Hamiltonians that did not generate the
Maxwell equations at all. This problem was analyzed by Hillery
and Mlodinow\cite{Hillery}, who showed that the problem was caused
by the use of an incorrect canonical description.

However, a previous macroscopic model of 
a nonlinear, dispersive medium
resulted in a quantum theory that has a consistent
Lagrangian \cite{Drum90}.  This was accomplished by
breaking the field up into narrow frequency-band components
and quantizing these separately.  The frequency dependent
linear susceptibilityÊis expanded up to second order in each
frequency band and this results in a local Lagrangian
in each band.
The present paper is intended to treat the microscopic basis of
the linear dispersion more precisely. Our model is the quantized 
version of  the classical  Drude-Lorentz \cite{Born}
single-electron oscillator model, in a linearized continuum limit.
Similar continuum models have been treated previously, generally in the minimal
coupling gauge.
 The present approach uses the multipolar gauge, which
eliminates complications due to Coulomb interactions and contact 
(${\bf A}^2$) terms. While these effects can be included in minimal coupling
theories, they result in considerable complexity. 
The present approach includes all these effects by virtue of the
transformation to a multipolar interaction, in which the coupling is to the
polarization rather than to the electronic momentum. This has the
advantage that inhomogeneous media or higher dimensions can be treated easily.

Some earlier dispersive models of a similar type are known, starting from the
phonon theory of Born and Huang \cite{BornH} and  
the exciton theory of Hopfield \cite{Hopfield}, which used  minimal
coupling. Hopfield considered the electromagnetic field coupled
to a polarization field which has a single resonant frequency.  He
diagonalized the resulting Hamiltonian and found that the resulting
elementary excitations, polaritons, have a dispersion relation 
consisting of two branches separated by a forbidden frequency band.
More recent work has revisited 
the Hopfield model, though not always with all the terms included 
in Hopfield's treatment \cite{Hutt91,Ho}.  Other treatments have
added losses to the model by coupling the polarization field to 
loss reservoirs \cite{Knoll,Hutt92},
still with just one resonant transition.  This has allowed the
examination of the Kramers-Kronig relations in a fully quantized
model.  Finally, nonlinear generalizations of of Hopfield's
model have also been studied \cite{Hillery}.  In these the
linear oscillators of Hopfield's medium have been replaced by
two-level atoms and the Holstein-Primakoff representation is
used to develop a systematic expansion of the polarization
of the medium which includes nonlinear effects.

These and other studies have elucidated the fundamental cause of
the two main types of dispersion.
The first, excitonic type is due to electronic transitions, and is typically resonant
in the ultra-violet. These  are due to relatively
tightly-bound electrons that are localized to atomic sites in insulators,
and somewhat delocalized in semiconductors.
The resulting classical dispersion is rather well described by the
harmonically-bound Drude-Lorenz model. The next type is due to phonon transitions,
mostly
in the infra-red. These excitations are less strongly localized, and have their own
dispersion relation in the Born-Huang model. The dispersion 
in the transparent region between these absorption bands
is caused by the
off-resonant, virtual transitions of these two main forms of excitation. In effect,
a photon propagates in a dielectric as a dressed particle, due to the virtual
transitions - and resulting polarization - that is induced in the solid.
To treat this completely, it is essential to include multiple exciton and
phonon resonances, in a full three-dimensional model.

The present paper examines this problem using an
approach in which the coupling to the dielectric is included
through a multipolar term \cite{GM,PZL}. This has the well-known advantages
that the use of atomic sum-rules
is not required, since the atomic transition probability
for off-resonant (high-frequency) transitions
is suppressed in this gauge \cite{Geltmann}. The new feature presented here
is the inclusion of
any number of atomic resonances - thus allowing us
to recover the Sellmeir\cite{Born} dispersion equation,
which is known to provide an excellent quantitative description
of real dispersive, transparent media. 
The use of multiple resonances is essential to the correct description
of the transparent region with low group velocity dispersion,
that occurs between the absorption bands in most known cases.
In addition,
there is a very straightforward procedure for diagonalizing the Hamiltonian -
even in the presence of an arbitrary number of resonant transitions
with different resonant frequencies.

In the homogeneous (plane-wave) case, we find the expansion
of the fields in terms of
the quantized modes has an exceptionally simple form,
in which only an electromagnetic group-velocity correction has to be
included relative to the usual non-dispersive theory.
When there is no longitudinal phonon/exciton dispersion,
this correction term is identical to that found in  previous 
phenomenological\cite{CDRS,DISPQ,Drum90}
 and single-resonance microscopic models\cite{BornH,Hopfield,Ho,Hutt91,Knoll,Hutt92}.
 As an example,
Blow, et. al. \cite{DISPQ} based their 
expansion of the fields on previous work (due to Loudon\cite {Loud63}) in 
which it was argued, from energy transport considerations, that 
the group velocity should appear in the expansion of the phonon
field in terms of quantized modes, when phonon dispersion is
present.  Blow, et. al. took this result for
phonons and used it for the case of an electromagnetic
field in a dispersive medium.  

Our work here provides a justification
for this quantization
procedure from a more fundamental point of view - a multiple-resonance
microscopic model.  
It is remarkable that  the simple inclusion of a group velocity
factor is sufficient to exactly diagonalize this complex interaction Hamiltonian.
The mathematical technique required to prove the results involves the use of Cauchy's
theorem to carry out the required sums over the different branches of the dispersion
relation. This is a significant improvement over previous approaches, since
in general there is no algebraic solution - even in principle - for the
polynomial expressions whose roots give the dispersion relations.
In addition, we find that when there is 
phonon or exciton dispersion\cite{Phdisp}, which is a realistic feature of
many dispersive media, this procedure is modified in a straightforward way.
Instead of the total group velocity, only the relevant electromagnetic
component is included in the expansion coefficients, suggesting that the
diagonalization is
 intimately related to the power transport processes in the solid.
  The
results apply to one, two or three-dimensional environments,
although only isotropic dielectrics are included here, for simplicity
in the derivations.

\section{One-dimensional model}

We start by considering a straight-forward one-dimensional model, without
the complicating features of transverse mode structures and longitudinal
phonon/exciton dispersion. This simple case illustrates the essential analytic
features of our model. It will be generalized to more realistic,
higher-dimensional dielectric structures, in later sections.

\subsection{One-dimensional Lagrangian}

We start by considering a set of 
dipole-coupled
electronic Lagrangians for oscillators with mass $m_\nu$,
charge $q_\nu$, and $N$ (renormalized) resonant discrete frequencies 
$\Omega_{\nu_1}...\ 
\Omega_{\nu_N}$.
These transitions may correspond physically to  different types of atom,
 to distinct transitions within each atom, or more generally
 to some many-body resonance. Thus,
\begin{equation}
{\cal L}_e=\sum\limits_\nu \left[ \frac {m_{\nu}} 2\left( \dot {{r} }%
_{\nu}^2-\Omega_{\nu}^2{r}_{\nu}^2\right) +\frac 1{\varepsilon _0}q_\nu{r}
_\nu
D(%
\overline{x}_\nu )\right] \ . 
\end{equation}
Here ${r} _\nu $ is the displacement of a
charged particle (typically, an electron) in a multi-polar
Drude-Lorentz model, from the corresponding center of charge
(typically, nuclear) position $\overline{x}%
_\nu $. Generally, these are all distinct atomic resonances. For
simplicity, the self-energy terms proportional to $P^2$, are incorporated
into the definition of the resonant frequency $\Omega_\nu $. The coupling
in multipolar gauge is
to the displacement field $D(\overline{x}_\nu )$ at the central position.
All Coulomb terms in this gauge
 are carried by the photon-exchange process\cite{PZL}.

Next, in order to develop a simple electromagnetic Lagrangian in the
multipolar gauge, we introduce a gauge field - the dual potential $\Lambda $%
, so that $D=\partial \Lambda /{\partial x}$ and $B=\mu  \partial 
\Lambda/ \partial t$. 
This simply corresponds to a canonical transformation 
to the multipolar gauge\cite{GM,PZL}
of the more
usual minimal coupling theory - and is simplified here by the
assumption that there are no free charges.
The discrete atomic positions are replaced by a continuum 
polarization field, together with the appropriate Lorentz-shifts in the
resonant frequencies $\nu$, due to local field corrections. This
simplification is used here because it is not really necessary
for us to consider the details of local-field corrections at this stage.
We will show that this approach is able to generate the identical
(low-frequency)
Sellmeir dispersion equation that is obtained in the usual
Drude-Lorentz theory, which does include local-field corrections.
Of course, the approximations involved are only valid for
carrier wavelengths much greater than the inter-atomic spacing.

The corresponding Lagrangian density that generates
the correct electromagnetic energy and Maxwell's equations in one dimension
is, for a one-dimensional wave-guide with effective
cross-section $A$:%
\begin{eqnarray}
{\cal L}/A&=&{1 \over 2\mu}{\dot 
{ \Lambda} }^2({ x})+
{1 \over \varepsilon _0}\left[
{ P} ({ x}) 
\partial_x { \Lambda} ({ x})
-
{1 \over 2}(\partial_x
{ \Lambda}({ x}) )^2 \right] \nonumber\\
&+& 
\sum\limits_{\nu=1}^N \left[ 
\dot { p}_{\nu}^2({ x})
-\Omega^2_{\nu}{ p}_{\nu}^2({ x})
\right]/({2 \varepsilon_0 g_\nu(x)})
. 
\end{eqnarray} 

Here $\rho _\nu (x)$ is the density of the oscillators with resonant
frequency $\nu $, and ${r} _\nu(x) $ is regarded as a continuum field, with
polarization $P(x)=\sum_\nu p_\nu(x)=\sum_\nu q_\nu\rho _\nu (x){r} _\nu
(x)$, in
order to allow us to use a continuum approximation in later calculations.

The coupling between the field and the polarization is entirely
included in one term $g_\nu (x)$, which effectively
combines all the relevant information about charge, density, and
dipole-moment:
\begin{equation}
g_\nu (x)=q_{\nu}^2\rho _\nu (x)/(m_\nu \varepsilon _0)\ . 
\label{gnu}
\end{equation}
Noting that $D(x)=\epsilon_0E(x)+P(x)$,
the electric field is given by $E(x)=[D(x)-P(x)]/\epsilon_0$, and
the resulting Maxwell-Bloch equations have the expected form of:%
\begin{eqnarray}
\partial _t^2{\Lambda }-c^2\partial _x^2\Lambda =-c^2\partial
_x P(x)\, 
\nonumber \\
\partial _t^2{p}_\nu +\Omega_{\nu}^2p_\nu =g_\nu \partial _x\Lambda \ . 
\end{eqnarray}

In order to quantize the system, we simply introduce commutators for the
canonical momenta and position, where:%
\begin{eqnarray}
\Pi(x) =\mu \dot{\Lambda }(x) = B(x)\ , 
\nonumber \\
\pi _\nu(x) = \dot{{p} }_\nu (x)/(\epsilon_0 g_{\nu}(x) )\ . 
\end{eqnarray}

The quantization rules are then the usual ones,
obtained on replacing Poisson brackets with commutators, except
with the atomic operators treated as a continuum field.
All position-type operators
of the same type of variable must commute amongst themselves,
as do all momentum-type operators. We list the commutators
involving cross-terms between the position and momentum operators,
and between the different types of variable, for reference in the next
sections. In order that the commutators will have a
familiar appearance, they are written in terms of the electric displacement
and magnetic fields, rather than the canonical variables and
their momenta. 
Since the electric field only differs from the displacement field
by the polarization terms (which commute with field operators
at equal times) there is no essential difference
between the electromagnetic commutators written with the
displacement field or the electric field.
The fact that the electric displacement is the derivative of a potential
is, of course, the origin of the differentiated
delta function in the commutator between the electric and magnetic fields,
which is also found in minimal coupling theory:

\begin{eqnarray}
({\rm I})\ \ \ \ \ \ \ \ \ \ \ \
\left[ \widehat{D }(x)\,,\,\widehat{B }(x^{\prime })\right] &=&i\hbar
\delta' (x-x^{\prime })/(A)\ , 
\nonumber \\
({\rm II})\ \ \ \ \ \ \ \ \ \ \ \
\left[ \widehat{p }_\nu (x)\,,\,\widehat{\pi }_{\nu ^{\prime
}}(x^{\prime })\right] &=&i\hbar \delta _{\nu \nu ^{\prime }}\delta
(x-x^{\prime })/A\ , 
\nonumber \\
({\rm III})\ \ \ \ \ \ \ \ \ \ \ \
\left[ \widehat{D}(x)\,,\,\widehat{p }_{\nu}(x^{\prime })\right]
&=&0 , 
\nonumber \\
({\rm IV})\ \ \ \ \ \ \ \ \ \ \ \
\left[ \widehat{B }(x)\,,\,\widehat{\pi }_{\nu }(x^{\prime })\right] &=&0 , 
\nonumber \\
({\rm V})\ \ \ \ \ \ \ \ \ \ \ \
\left[ \widehat{D }(x)\,,\,\widehat{\pi }_{\nu }(x^{\prime })\right]
&=&0
, 
\nonumber \\
({\rm VI})\ \ \ \ \ \ \ \ \ \ \ \
\left[ \widehat{B }(x)\,,\,\widehat{p }_{\nu }(x^{\prime })\right] &=&0 , 
\end{eqnarray}

it is important to notice here that the commutators are essentially identical
to those for the corresponding free fields\cite{Lee,Lois} and oscillators. This is
a consequence of the fact that the couplings do not involve time-derivatives.
It also demonstrates that the present theory is canonically equivalent to
other techniques involving minimal coupling\cite{GM,PZL}.
 One apparent difference is in the
commutator between the displacement field and the  momentum
$\widehat{\pi }_{\nu }$, which replaces a commutator between the electric field
and a gauge-dependent canonical momentum in minimal-coupling theory. Since
the electric field and displacement field differ by a term that doesn't
commute with the canonical momentum, it might seem that this introduces
a difference. The explanation is due to the well-known fact that the
canonical momentum in this case is just the usual mechanical momentum,
and therefore differs from the minimal coupling momentum -
which includes a term proportional to the vector potential $A$. These
two effects cancel each other, so the overall commutators are unchanged.

The final Hamiltonian density, equal to the linear energy density of the
coupled system, has a rather straightforward expression
in which there are no explicit Coulomb interaction terms. This is
a typical property of multipolar interaction Hamiltonians. Effectively,
the Coulomb couplings are included partly in the
oscillator self-frequency terms (i.e., in $\nu_i$),
and partly in the photon-exchange dynamics that are
implicit in the Hamiltonian. The result is that:
\begin{eqnarray}
{\cal H}/A&=&
 {1\over 2\varepsilon _0} \widehat{{ D} }^2({ x})  
+{1\over 2 \mu } \widehat{{ B} }^2({ x})  
-
{1\over \varepsilon _0}\widehat{{ D} }({ x}) \widehat{{ P} }({ x}) 
 \\ &+& 
{1\over 2\varepsilon _0 g_\nu(x) }\sum\limits_{\nu =1}^N 
\left[ \varepsilon^2 _0 g_\nu(x)^2 
\widehat{ \pi }_{\nu}^2({ x}) +
\Omega^2_{\nu}\widehat{{ p} }_\nu^2 ({ x})
 \right]
 \label{honed} 
\end{eqnarray}

All these equations hold for an arbitrary spatial distribution 
$\rho(x)$ of
the polarizable atoms. If $\rho(x)$ is a sum of delta-function
terms, a discrete lattice model is obtained; it
is an unusual waveguide in which the atoms are all lined up
in a one-dimensional row, but not an impossible one, with atom-trapping
techniques.
For simplicity, we do not treat this type of model here. Instead, we will
focus on the uniform dielectric limit, in which all the local-field corrections
are included via the relevant Lorentz shifts of the oscillator frequencies,
to give a more tractable case.

\subsection{Mode Structure}

We now specialize to the case of a
continuum model with a uniform distribution, in order to
find the dispersion relations for plane-waves.
We introduce a mode structure by finding the eigenmodes of the equation of
motion. Suppose the solutions to Maxwell's equations have the form:%
\begin{equation}
\stackrel{\rightarrow }{\lambda }(t,x)=\left( 
\begin{array}{c}
\lambda (t,x) \\ 
p_\nu (t,x)
\end{array}
\right) =\left( 
\begin{array}{c}
\widetilde{\lambda } \\ \widetilde{p}_\nu 
\end{array}
\right) e^{ikx-i\omega t}\ . 
\end{equation}

It follows that these satisfy the resulting equations in the form:%
\begin{eqnarray}
\left( -\omega ^2+c^2k^2\right) \widetilde{\lambda }&=&-ikc^2\sum\limits_{\nu
^{\prime }}\widetilde{p}_{\nu ^{\prime }}\nonumber \\  
(-\omega ^2+\Omega_{\nu}^2)\widetilde{p}_\nu &=&ikg_\nu \widetilde{\lambda }\
. 
\end{eqnarray}

Solving for $\widetilde{p}_\nu $, we find that:%
\begin{equation}
(\Omega_{\nu}^2-\omega ^2)(c^2k^2-\omega ^2)\widetilde{p}_\nu =c^2k^2g_\nu
\sum\limits_{\nu ^{\prime }}\widetilde{p}_{\nu ^{\prime }}\ , 
\end{equation}
or, summing over all the oscillators  and introducing
$\widetilde p=\sum_\nu \widetilde p_\nu $, we find:
\begin{equation}
(c^2k^2-\omega ^2)\widetilde p=
\widetilde pc^2k^2\sum\limits_\nu \frac{g_\nu }{\Omega_{\nu}^2-\omega ^2}\
. 
\end{equation}

Eliminating the polarization field $\widetilde p$ leads to the eigenvalue
equation:%
\begin{equation}
\label{eigen}
{\omega ^2}= \frac{c^2k^2}{n^2(\omega )}=c^2k^2
\left[1-\sum\limits_\nu \frac{g_\nu }{\Omega_{\nu}^2-\omega ^2}\right] \, . 
\end{equation}

We find a band-structure in which there are
typically ($N+1$) positive roots $\omega_{\mu}(k)$ 
with $\mu = 0,1,.. N$
to the solution. 
To look at this differently,
we can solve for the wave-number $k$, 
at any given frequency $\omega$: 
\begin{equation}
k=\pm \left[ \frac{\omega ^2/c^2}{1-\sum\limits_\nu g_\nu
/(\Omega_{\nu}^2-\omega ^2)%
}\right] ^{1/2}=\pm k(\omega )\ . 
\end{equation}
This solution is unique for any given modal frequency, but has forbidden
regions at $\omega \simeq \Omega_{\nu} $, where $k^2\rightarrow -\infty $. 
This indicates a resonance, or absorption band. In the transmission bands,
there
is a unique refractive index $n(\omega)$, and hence a well-defined
permittivity,
$\varepsilon(\omega) = n^2(\omega)\varepsilon_0$.

It should be realized that the
dispersion relation is not completely identical to the usual 
classical Sellmeir
expansion, although it is very similar to it. The Sellmeir expansion
is:%
\begin{equation}
n^2(\omega )=1+\sum\limits_\mu \frac{\tilde g_\mu
}{\tilde\Omega_{\nu}^2-\omega
^2}\ . 
\end{equation}
This agrees with the functional form of the present result to lowest order
in $g_\nu $,
if we assume that $g_\nu = \tilde g_\mu$. 
The difference is simply due to different approaches to
treating local-field corrections in the continuum approximation. If a
precise local-field theory is required, then the assumption of a homogeneous
polarization field must be replaced by a lattice model. 
In fact,
the usual Drude-Lorentz model is not typically
obtained from a coupled Lagrangian, so
it cannot be readily quantized directly.
 Rather, it is obtained from an approximate
theory in which the local (microscopic)
$E$ field plays the role that the $D$ field does in the
present theory\cite{Born}. From microscopic considerations, both forms needs
to have local-field
corrections included near an absorption band, in order to give an accurate
comparison with a three-dimensional crystal lattice, from first principles.

When this is done in the Drude-Lorentz model,
all the low-frequency resonances are shifted by amounts known
as the Lorentz shift. With these shifts included, the Sellmeir
expansion is obtained, with local-field corrections included.
However, the number of poles in the rational
function representation derived here is finite, just as in the 
classical Drude-Lorentz theory.
Accordingly, it is always possible to re-express our dispersion
relation exactly in the Sellmeir rational-function form,
using partial fraction expansions, just as
in the Drude-Lorentz derivation of the Sellmeir equation. In this
case, the frequencies and couplings $\tilde g_\mu$ appearing
in the final Sellmeir formula are not identical with the
original frequencies in the Lagrangian; these shifts of course,
can be regarded as evidence of the photonic coupling
between the oscillators in our model.

An important, 
and experimentally well-tested
property of the Sellmeir equation is that the refractive index approaches
unity at high enough frequencies, while at low frequencies it approaches
a constant  value greater than one:
\begin{equation}
\lim_{\omega \rightarrow  0}n^2(\omega )=1+\sum\limits_\mu \frac{\tilde g_\mu
}{\tilde\Omega_{\nu}^2}\ . 
\end{equation}
Our dispersion relation from the multipolar Hamiltonian has
a similar behavior, except that
the algebraic form is slightly different at low frequencies:

\begin{equation}
\lim_{\omega \rightarrow  0}n^2(\omega )=
\left[1-\sum\limits_\nu \frac{ g_\nu }{\Omega_{\nu}^2}\right]^{-1}\ . 
\end{equation}

Clearly, one form can always be transformed into
another, given the obvious restrictions
on having distinct roots, with $\sum_\nu { g_\nu }/{\Omega_{\nu}^2} < 1$.
We note that this equivalence does not apply in all other
models of the dispersion relations, which may have inequivalent
analytic properties - leading to a different limiting
behavior at high and low frequencies.
Provided the general analytic properties are equivalent,
the partial-fraction procedure 
to transform one form into the other is not
required in most cases. We shall
demonstrate that only the refractive index and group velocity
are needed to obtain a complete quantum theoretic
description of the modes. This
information is readily available from the usual Sellmeir
parameters that are experimentally known for many transparent
materials. A typical dispersion relation for a solid
with both low and high-frequency resonances, would
show multiple transmission and absorption
bands - with three distinct branches to the dispersion curve -
and a region of relatively low group-velocity dispersion
between the absorption bands. This is the origin of the
well-known zero-dispersion point (at around $\lambda = 1.5 \mu m$
in fused silica), which plays a prominent role in optical
communications systems.
 
\section{One-dimensional mode operators}

Having derived the modal solutions, we now wish to expand the fields of the
theory in terms of annihilation and creation operators. 
We develop this expansion in two stages. First, we will consider the
conditions on the mode expansion which preserve the correct commutation
relations
for the original canonical fields. Then, we show that a mode expansion
defined this way does lead to a diagonal Hamiltonian form, when the
Hamiltonian
is re-expressed in terms of the mode operators. That is, our goal is to
find operators $\widehat{a}_{\mu}(k)$
which  have the function of diagonalizing the Hamiltonian, giving
the final structure of:
\begin{equation}
H=\sum\limits_{\mu=0}^{N}\int \hbar \omega _{\mu}(k)\widehat{a}_{\mu}^{\dagger
}(t,k)%
\widehat{a}_{\mu}(t,k)\,dk\ . 
\end{equation}

\subsection{Commutation properties}

Clearly, since the above expansion must lead to the same eigenfrequencies as
the original Maxwell equations, we should
define mode operators $\widehat{a}_{\mu}$ relative to each branch
of the dispersion relation (with,
for the sake of simplicity, $N+1$ distinct branches) so that:%
\begin{equation}
\widehat{\Lambda }(t,x)=\sum\limits_{\mu=0}^{N}\int dk\left[ \Lambda
_{\mu}(k)%
\widehat{a}_{\mu}(t,k)e^{ikx}+h.c.\right] \ . 
\end{equation}
Here $\omega _{\mu}(k)$ is the inverse of $k(\omega )$, for the $\mu$-th
branch.
The summation is over the discrete branches in the dispersion relation, each
of which correspond to a different `particle' type. 
The time-dependence of the mode operators 
in the Heisenberg picture 
- given the desired Hamiltonian
structure - must be:
\begin{equation}
\widehat{a}_{\mu}(t,k) = \widehat{a}_{\mu}(k)e^{-i\omega _{\mu}(k)t} \ . 
\end{equation}

These combined field-polarization excitations can be termed polaritons.
We will suppose that the 
$\widehat{a}_{\mu}(k)$ are chosen so that:%
\begin{equation}
\left[ \widehat{a}_{\mu}(k),\widehat{a}_{\mu'}^{\dagger }(k^{\prime })\right]
=\delta
_{\mu\mu'}\delta (k-k^{\prime })\ . 
\end{equation}

Similarly, the momentum field can be expanded as:%
\begin{equation}
\widehat{\Pi }(t,x)=\sum\limits_{\mu=0}^{N}\int dk\left[ \Pi
_{\mu}(k)\widehat{a}%
_{\mu}(k)e^{ikx-i\omega _{\mu}(k)t}+h.c.\right] \ . 
\end{equation}

The requirement of commutation relations means that we must have (at equal
times):%
\begin{equation}
\left[ \widehat{\Lambda }(x)\,,\,\widehat{\Pi }(x^{\prime })\right] =i\hbar
\delta (x-x^{\prime })/A\ =\sum\limits_{\mu=0}^{N}\int dk\left[ \Lambda
_{\mu}(k)\Pi _{\mu}^{*}(k)e^{ik(x-x^{\prime })}-h.c.\right] \ . 
\end{equation}

This implies the relationship that, in order to preserve commutation
relations,%
\begin{equation}
\sum\limits_{\mu=0}^{N}\Lambda _{\mu}(k)\Pi _{\mu}^{*}(k)=\frac{i\hbar }{4\pi
A}\ . 
\end{equation}

The Lagrangian density for this model implies that $\Pi =\mu\partial_t 
\Lambda $.
With the assumed time dependence of the annihilation operators, 
we can also write $\Pi _{\mu}(k)$ in the form of:%
\begin{equation}
\Pi _{\mu}(k)=-i\omega _{\mu}(k)\mu \Lambda _{\mu}(k)\ . 
\end{equation}

The above result therefore reduces to
an equation for the expansion coefficients $\Lambda _{\mu}(k)$,
in the form of:%
\begin{equation}
\sum\limits_{\mu=0}^{N}\omega _{\mu}(k)\Lambda _{\mu}^2(k)=\frac \hbar {4\pi
A\mu } 
\end{equation}

Next, we wish to obtain an expression for $\Lambda _{\mu}(k)$, which is
unknown
at this stage. It is no restriction to choose $\Lambda _{\mu}(k)$ to be real.
Therefore, we can always choose an
(unknown) function $v_{\mu}(k)$ so that, in analogy to
the standard vacuum expansion,%
\begin{equation}
\Lambda _{\mu}(k)=\left[ \frac{\hbar v_{\mu}(k)\varepsilon _{\mu}(k)}{4\pi
Ak}\right]
^{1/2}\ . 
\end{equation}

If $v_{\mu}(k)=c$ and $\varepsilon _{\mu}(k)=\varepsilon _0$,
this
reduces to the accepted vacuum field expansion. More generally, we define $%
\varepsilon _{\mu}(k)=k^2/[\mu \omega _{\mu}^2(k)]$ as the effective
permittivity of
the $\mu$-th branch. We will show later that $v_{\mu}(k)$ must be interpreted
as
the group velocity, just as in an earlier narrow-band analysis
of this problem, using effective Lagrangian arguments\cite{CDRS,Drum90}.

In order to demonstrate this, we first substitute the above expression
for $\Lambda _{\mu}(k)$ into the equation for
 the consistency of the field and
mode-operator commutation relations (i.e. for $[\widehat{\Lambda
},\widehat{\Pi }%
] $ and $[\widehat{a},\widehat{a}^{\dagger }]$). This leads to
the very simple result that:%
\begin{equation}
({\rm I})\ \ \ \ \ \ \ \ \ \ \ \ \sum\limits_{\mu=0}^{N}\frac{kv_{\mu}(k)}{%
\omega _{\mu}(k)}=1\, . 
\end{equation}

As explained above, 
we have to determine a function $v_{\mu}(k)$ which satisfies this condition,
and we intend to demonstrate that the choice of $v_{\mu}(k)$
as the group-velocity of the relevant polariton branch is sufficient;
no other correction factors are needed in this simple model.
In order to verify this,
 we can  differentiate both sides of Eq. (\ref{eigen}) with respect to $k$. 
This gives a group velocity of:
\begin{equation}\label{vg1}
v_{\mu}(k) =
\frac{d\omega_{\mu}(k)}{dk}=\frac{\omega_{\mu}(k)}{k}\left(1+\sum_{\nu}
\frac{k^{2}c^{2}g_{\nu}}
{(\Omega_\nu^{2}-\omega_{\mu}^{2}(k))^{2}}\right)^{-1} \, ,
\end{equation}
which is the function we propose to use in the mode expansion.

However,  it is clear that the mode function expansion of $%
\widehat{p}_\nu$ and $\widehat{\pi }_\nu $ are also needed,
for a complete demonstration of consistency. Using
Maxwell's equations, we note that for a Fourier component of $p _\nu $
at frequency $\omega $, wave-vector $k$, we must have:%
\begin{equation}
\widetilde{p }_\nu = 
\frac{ik g_{\nu}
\widetilde{\lambda }}{\Omega_{\nu}^2-\omega ^2}\ . 
\end{equation}
Thus, if we expand $\widehat{p }_\nu $ as:%
\begin{equation}
\widehat{p }_\nu =\sum\limits_{\mu=0}^{N}\int dk \left[ p _{\mu}^\nu
(k)%
\widehat{a}_{\mu}(k)e^{ikx-i\omega _{\mu}(k)t}+h.c.\right] \ , 
\end{equation}
it follows that the expansion coefficient for in the $j$-th frequency band
is:%
\begin{equation}
p _{\mu}^\nu (k)=\frac{ikg_{\nu}\Lambda _{\mu}(k)}
{(\Omega_{\nu}^2-\omega^2_{\mu}(k))}%
\ . 
\end{equation}

Similarly, if the canonical momentum for the atomic polarization field is:%
\begin{equation}
\widehat{\pi }_\nu (t,x)=\sum\limits_{\mu=0}^{N}\int dk\left[ \pi _{\mu}^\nu
(k)%
\widehat{a}_{\mu}(k)e^{ikx-i\omega _{\mu}(k)t}+h.c.\right] \ , 
\end{equation}
then the corresponding expansion coefficient is:%
\begin{equation}
\pi _{\mu}^\nu (k)=\frac{ k\omega _{\mu}(k)\Lambda _{\mu}(k)}{\varepsilon
_0(\Omega_\nu
^2-\omega _{\mu}^2(k))}\ . 
\end{equation}

For these operators to have the correct equal-time commutators, 
 the different oscillator position operators
 $\widehat{p }_\nu$ must commute amongst themselves
 at equal times, as must the
 different momentum operators $\widehat{\pi }_\nu $. This is trivial
 from the form of the 
 mode operator expansion. However, the commutation
 relations (II) between the position and momentum operators are non-trivial.
The relevant commutation
conditions are therefore:%
\begin{equation}
\left[ \widehat{p }_\nu (x)\,,\,\widehat{\pi }_{\nu ^{\prime
}}(x^{\prime })\right] =i\hbar \delta _{\nu \nu ^{\prime }}\delta
(x-x^{\prime })/A\ =\sum\limits_{\mu=0}^{N}\int dk\left[ p _{\mu}^\nu
(k)\pi _{\mu}^{*\nu ^{\prime }}(k)e^{ik(x-x^{\prime })}-h.c.\right] \ . 
\end{equation}
This in turn implies that:%
\begin{equation}
\sum\limits_{\mu=0}^{N}p _{\mu}^\nu (k)\pi _{\mu}^{*\nu ^{\prime }}(k)=\frac{%
i\hbar }{4\pi A}\delta _{\nu \nu ^{\prime }}\ . 
\end{equation}

Expanding the coefficients gives the new equation:%
\begin{equation}
\sum\limits_{\mu=0}^{N}\frac{\omega _{\mu}(k)k^2g_{\nu}\Lambda _{\mu}^2(k)
}{\varepsilon _0(\Omega_{\nu}^2-\omega_{\mu}^2(k))
(\Omega_{\nu ^{\prime}}^ 2-\omega^2_{\mu}(k))}=%
\frac \hbar {4\pi A}\delta _{\nu \nu ^{\prime }}\ . 
\end{equation}
However, since $\varepsilon _{\mu}(k)=c^2k^2\varepsilon _0/[ \omega
_{\mu}(k)]$, and hence%
\begin{equation}
\Lambda _{\mu}^2(k)=\frac{\hbar c^2\varepsilon _0kv_{\mu}(k)}{4\pi  A\omega
_{\mu}^2(k)}\ , 
\end{equation}
this simplifies to the form:%
\begin{equation}
({\rm II})\ \ \ \ \ \ \ \ \sum\limits_{\mu=0}^{N}\frac{c^2k^3v_{\mu}(k)
g_{\nu}}{\omega _{\mu}(k)(\omega _{\mu}^2(k)-\Omega_{\nu}^2)
(\omega_{\mu}^2(k)-\Omega_{\nu ^{\prime}}^ 2)}=\delta _{\nu \nu ^{\prime }}\ . 
\end{equation}

Finally, to ensure that 
there are correct field-atom commutators, we must 
satisfy the commutators III-VI. For these cross-terms between the
oscillators and field variables, we find that conditions (III)
and (IV), involving commutators between the field and the 
particle position (or the field momentum and particle momentum)
are automatically satisfied. 
This occurs for the same reason that commutators like 
$[\widehat{\Lambda }(x),\widehat{\Lambda }(x')]$ 
or $[\widehat{\Pi }(x),\widehat{\Pi }(x')]$ must equal zero
in our mode expansion. In all these cases involving pairs of 
canonical position-type operators or pairs of momentum-type
operators, the commutator reduces to an integral over $k$,
which is an odd function of $k$ - and hence vanishes when integrated over
all positive and negative k-values.

This leaves the 
requirements (V) and (VI), which are that
$\widehat{\Lambda }$ and $\widehat{\pi }_\nu$ must commute at
equal times ,
as well as $\widehat{\Pi }$ and $\widehat{\delta}_\nu$. These
two requirements {\it both} imply the same restriction
on the expansion coefficients, and hence on $v_{\mu}(k)$,
which is that
 for all $k$ and $\nu $ we must have the conditions:%
\begin{equation}
({\rm V},{\rm VI})\ \ \ \ \ \ \ \ \ \ \ \ \sum\limits_{\mu=0}^{N}\frac{
kv_{\mu}(k)}{\omega _{\mu}^2(k)(\omega _{\mu}^2(k)-\Omega_{\nu}^2)}=0\ .
 \end{equation}

Despite the complex nature of each of these conditions - which involve
sums over all the roots of the dispersion equation, and must be satisfied
for all the resonant frequencies $\nu$, as well all momenta $k$ -
we will show that each of these sums can be analytically evaluated
without requiring an algebraic solution for the roots. 
 
\subsection{Conditions on expansion coefficients}

From the previous
results, we have shown that the condition on the expansion coefficient of mode
operators is that we must find a function $v_{\mu}(k)$, such that:%
\begin{equation}
S^{({\rm I})}=\sum\limits_{\mu=0}^{N}\frac{kv_{\mu}(k)}{\omega _{\mu}(k)}=1\ , 
\end{equation}
together with an orthogonality condition:%
\begin{equation}
S_{\nu \nu ^{\prime }}^{({\rm II})}=\sum\limits_{\mu=0}^{N}\frac{%
c^2k^3v_{\mu}(k)g_{\nu ^{\prime }}}{\omega _{\mu}(k)(\omega
_{\mu}^2(k)-\Omega_{\nu}^2)(\omega
_{\mu}^2(k)-\Omega_{\nu^{\prime}}^{2})}=\delta _{\nu \nu ^{\prime }}\ . 
\end{equation}

In addition, to ensure commutation between the particle and electromagnetic
fields, we should impose the condition:%
\begin{equation}
S_\nu ^{({\rm III})}=\sum\limits_{\mu=0}^{N}\frac{kv_{\mu}(k)}
{\omega _{\mu}^2(k)(\omega _{\mu}^2(k)-\Omega_{\nu}^2)}=0\ . 
\end{equation}

Earlier work\cite{Drum90} on more phenomenological 
narrow-band quantum models of dispersion
led to the conclusion that, for an expansion of modes to be consistent
with both Maxwell's equations and the known dispersive energy,
it is necessary to choose $v_{\mu}(k)$ equal to the group velocity.
Thus, the use of $v_{\mu}(k)=\partial \omega _{\mu}(k)/\partial k$ is
an obvious choice, but it is necessary to demonstrate that
this still leads to a complete set of consistent
commutation relations.

\subsection{Single Oscillator Case}

As an example,  we consider the single-oscillator case, where the dispersion
relation 
can be treated algebraically as the solution
of a quadratic equation. In this case
the refractive index is given by:%
\begin{equation}
n(\omega ) ^2=[1-\frac g{\Omega_{\nu}^2-\omega ^2}]^{-1}\ . 
\end{equation}

In order to show how the Sellmeir form can be regained,
we define a new frequency $\tilde\Omega_\nu^2 = \Omega_{\nu}^2 -g$. 
As long  as $\Omega_{\nu}^2  > g$,
the above equation is equivalent to
a Sellmeir type of dispersion relation,
which is simply:

\begin{equation}
n(\omega )^2=1+\frac g{\tilde\Omega_{\nu}^2-\omega ^2}\ . 
\end{equation}

Either equation leads to a quadratic for $\omega ^2$, having the form:%
\begin{equation}
\omega ^4-\omega ^2(c^2k^2+\Omega_{\nu}^2)+\Omega_{\nu}^2-g=0\ . 
\end{equation}
The resonant frequencies at any given wavenumber $k$, are then:%
\begin{equation}
\omega ^2=\frac 12\left( c^2k^2+\Omega_{\nu}^2\pm \sqrt{(c^2k^2+\nu
^2)^2-4c^2k^2(\Omega_{\nu}^2-g)}\right) \ . 
\end{equation}
We now identify $v_{\mu}(k)$
with the group-velocity  on each of the two branches,
by taking derivatives with respect to $k$. Thus, assuming $\Omega_{\nu}^2>g$
(to
have distinct branches):%
\begin{equation}
\frac{kv_{\pm }(k)}{\omega _{\pm }(k)}=\frac 12\left( \frac{c^2k^2}{\omega ^2%
}\right) \left( 1\pm \frac{c^2k^2-\Omega_{\nu}^2+2g}\Delta \right) \ , 
\end{equation}
where the quantity $\Delta $ is given by%
\begin{equation}
\Delta =\sqrt{(c^2k^2+\Omega_{\nu}^2)^2-4c^2k^2(\Omega_{\nu}^2-g)}\ . 
\end{equation}
Clearly it is necessary to have $\Omega_{\nu}^2>g$ in order to have distinct
real
branches to the dispersion relation, each with positive frequency $\omega $.
This is precisely the condition under which the Sellmeir expansion
is valid, as one might have expected.

Summing over the two branches, we note that (defining $\widetilde{K}%
=c^2k^2+\Omega_{\nu}^2$):%
\begin{equation}
\sum_{\pm}\frac{kv_{\pm }(k)}{\omega _{\pm }(k)}=\frac 14%
\sum_{\pm }\left( 1\pm \frac{\widetilde{K}+2(g-\Omega_{\nu}^2)}\Delta \right) 
\frac{\widetilde{K}\mp \Delta }{\Omega_{\nu}^2-g}\ . 
\end{equation}
On taking the sum, this reduces to the required result of:%
\begin{equation}
S^{({\rm I})}=\sum\limits_{\pm}\frac{kv_{\pm }(k)}{\omega _{\pm }(k)}%
=1\ . 
\end{equation}

This indicates that the use of group-velocity expansion coefficients appears
correct in this case,
although we have only calculated the simplest of the commutators.
However, this algebraic technique is rather clumsy to use for the other
identities. Even worse, it is not able to be used at all
for an arbitrary number of branches; in these more general cases there is 
no closed form expression for the solution to the dispersion equation.

\section{Analytic properties of dispersion relations}

For the other, more complex, 
commutation relation identities - or more
oscillators - it is preferable to use techniques from complex
function theory,
which transform the sums over roots of the dispersion relation
to complex contour integrals of related meromorphic functions.
The dispersion relations considered here have the general structure of:%
\begin{equation}
\frac{\omega ^2}{c^2k^2}=1-\sum\limits_\nu \frac{g_\nu }{\Omega_{\nu}^2-\omega
^2}\
. 
\end{equation}
This can be written in the form of roots of a polynomial in $z=\omega ^2$,
so that:%
\begin{equation}
{\kappa}a(z_{\mu})=b(z_{\mu})\ , 
\end{equation}
where $z_{\mu}=\omega _{\mu}^2$, ${\kappa}=c^2k^2$ and:%
\begin{equation}
a(z)=\prod\limits_\nu (\Omega_{\nu}^2-z)-\sum\limits_{\nu ^{\prime }}g_{\nu
^{\prime
}}\prod\limits_{\nu \neq \nu ^{\prime }}(\Omega_{\nu^{\prime}}^{2}-z)\ , 
\end{equation}
\begin{equation}
b(z)=z\prod\limits_\nu (\Omega_{\nu}^2-z)\ . 
\end{equation}
Next, 
in order to test the assumption that the expansion coefficients involve
group velocities, we must
consider the slope of the dispersion relations:%
\begin{equation}
{\kappa}a^{\prime }(z)+\frac{\partial {\kappa}}{\partial z}a(z)=b^{\prime
}(z)\
. 
\end{equation}
Hence,%
\begin{equation}
v_{\mu}(k) = {\partial \omega_{\mu}(k) \over \partial k}
={c^2 k a(z) 
\over \omega_{\mu}(k)(b^{\prime }(z)-{\kappa}a^{\prime })}\ . 
\end{equation}

A simpler way to write this expression - entirely equivalent to the above
definition - is Eq. (\ref{vg1}).
While this form is more transparent, the above expression is 
a rational function, which is amenable to analysis using Cauchy's theorem.

\subsection{Condition I}

The sum $S^{({\rm I})}$ has the form:%
\begin{equation}
S^{({\rm I})}=\sum\limits_{\mu}\frac{kv_{\mu}(k)}{\omega
_{\mu}(k)}=\sum\limits_{\mu}\left[ 
\frac{{\kappa}a(z)}{z\left( b^{\prime }(z)-{\kappa}a^{\prime }(z)\right)
}\right]
_{z=z_{\mu}({\kappa})}\ . 
\end{equation}

Next, consider the complex function, defined for the analytic continuation
of $z$ into complex values:%
\begin{equation}
f^{({\rm I})}(z)=\frac{{\kappa}a(z)}{z\left[ b(z)-{\kappa}a(z)\right] }\ . 
\end{equation}
This generally has ($N+2$) poles; and has the property that $\lim
\limits_{|z|\rightarrow \infty }f^{({\rm I})}(z)\sim 1/z^2$. For example, in
the trivial case of no oscillators ($N=0$), we find that:%
\begin{equation}
f^{({\rm I})}(z)=\frac {\kappa}{z(z-{\kappa})}\ . 
\end{equation}
In this case, the identity (I) is satisfied trivially, since it reduces to
Res$\left[ f^{({\rm I})}(z=c^2 k^2)\right] =1$. The sum of
residues of $f^{({\rm I})}(z)$ is zero in this case, which must be true in
general for a meromorphic function behaving as $f^{({\rm I})}(z)\sim 1/z^2$
as $|z|\rightarrow \infty $. As usual in complex function theory of the
inverse variable ($u=1/z$), a contour integral around all the finite poles
turns into a contour integral around zero poles at infinity, and hence must
equal zero. Thus, we have the general result that:%
\begin{equation}
0=\sum {\rm Res}\left[ f^{({\rm I})}(z)\right]
=-1+\sum\limits_{\mu}^{N+1}\left[ 
\frac{{\kappa}a(z_{\mu})}{z_{\mu}\left( b^{\prime }(z_{\mu})-{\kappa}a^{\prime
}(z_{\mu})\right) }\right]
\ . 
\end{equation}
However, this is precisely condition (I), for the $N$-oscillator case, since:%
\begin{equation}
S^{({\rm I})}=\sum\limits_{\mu}^{N+1}\left[
\frac{{\kappa}a(z_{\mu})}{z_{\mu}\left( b^{\prime
}(z_{\mu})-{\kappa}a^{\prime }(z_{\mu})\right) }\right]
=\sum\limits_{\mu}^{N+1}\frac{kv_{\mu}(k)}{%
\omega _{\mu}(k)}=1\ . 
\end{equation}
Thus, the use of complex function theory shows that (I) is always satisfied,
provided there are ($N+1$) distinct roots.

\subsection{Condition II}

Similarly, we can prove the other relations. For example, to prove relation
(II) we consider:%
\begin{equation}
f_{\nu \nu ^{\prime }}^{({\rm II})}(z)=\frac{{\kappa}g_{\nu ^{\prime
}}f^{({\rm
I}%
)}(z)}{(z-\Omega_{\nu}^2)(z-\Omega_{\nu ^{\prime}}^ 2)}\ . 
\end{equation}

Summing the residues of this function, and noting that $\lim
\limits_{z\rightarrow 0}f^{({\rm I})}(z)=-1/z$, we find (for $\nu \neq \nu
^{\prime }$):%
\begin{equation}
0=S_{\nu \nu ^{\prime }}^{({\rm II})}-\frac{{\kappa}g_{\nu ^{\prime
}}}{\Omega_{\nu}^2
\Omega_{\nu^{\prime}}^ 2}+\frac{{\kappa}g_{\nu ^{\prime }}f^{({\rm
I})}(\Omega_{\nu}^2)}{(\Omega_{\nu}^2-
\Omega_{\nu^{\prime}}^ 2)}+\frac{{\kappa}g_{\nu ^{\prime }}f^{({\rm I})}
(\Omega_{\nu^{\prime}}^{2})}{(
\Omega_{\nu^{\prime}}^ 2-\Omega_{\nu}^2)}\ . 
\end{equation}
Examining the RHS of the required identity, we must obtain the value of $f^{(%
{\rm I})}(\Omega_{\nu}^2)$, evaluated at each resonance:%
\begin{equation}
f^{({\rm I})}(\Omega_{\nu}^2)=\frac{{\kappa}
a(\Omega_{\nu}^2)}{\Omega_{\nu}^2\left[ b(\Omega_{\nu}^2)-{\kappa}a(\Omega_\nu
^2)\right] }\ . 
\end{equation}

However, $b(\Omega_{\nu}^2)=0$ at each resonance, so that $f^{({\rm
I})}(\Omega_{\nu}
^2)=-1/\Omega_{\nu}^2$. Hence, the RHS of the required identity becomes:%
\begin{equation}
S_{\nu \nu ^{\prime }}^{({\rm II})}={\kappa}g_{\nu ^{\prime }}\left[ \frac
1{\Omega_\nu
^2\Omega_{\nu ^{\prime}}^ 2}+\frac 1{\Omega_{\nu}^2(\Omega_{\nu}^2-\Omega_{\nu
^{\prime }}^2)}
+\frac
1{\Omega_{\nu
^{\prime}}^ 2(\Omega_{\nu ^{\prime}}^ 2-\Omega_{\nu}^2)}\right]  = 0\ . 
\end{equation}

In the case that $\nu =\nu ^{\prime }$, a double pole is found, so the
residue is obtained on differentiating $f^{({\rm I})}(z)$. We can perform
this operation most simply in the neighborhood of the double root at
$z=\Omega_\nu
^2$, by using the definition of $f^{({\rm I})}(z)$ to show that:%
\begin{equation}
f^{({\rm I})}(z)=-\frac 1z+\frac 1{z-{\kappa}\left( 1-\sum\limits_\nu g_\nu
/(\Omega_\nu
^2-z)\right) }\ . 
\end{equation}
Thus, as $z\rightarrow \Omega_{\nu}^2$, we find the second term is dominated
by
the
pole in the denominator:%
\begin{equation}
\lim \limits_{z\rightarrow \Omega_{\nu}^2}f^{({\rm I})}(z)=-\frac
1z+\frac{\Omega_{\nu}^2-z}{%
{\kappa}g_\nu }\ . 
\end{equation}
Hence, on differentiating to obtain the residue,%
\begin{equation}
\frac \partial {\partial z}\left. f^{({\rm I})}(z)\right|
_{z=\Omega_{\nu}^2}=-\frac 
1{{\kappa}g_\nu }+\frac 1{\Omega_\nu ^4}\ . 
\end{equation}
This is sufficient to complete the proof of the second relation, which is%
\begin{equation}
S_{\nu \nu ^{\prime }}^{({\rm II})}
=\sum {\rm Res}\left[ f_{\nu \nu ^{\prime }}^{({\rm II})}(z)\right]
+\delta _{\nu \nu ^{\prime }}= \delta _{\nu \nu ^{\prime }} \ . 
\end{equation}

\subsection{Condition III-VI}

As shown previously, the conditions (III) -(IV) are straightforward
consequences of the general type of expansion chosen here,
while conditions (V) and (VI) reduce to an identical summation identity.
To obtain this last identity,
we can now introduce a third analytic function,%
\begin{equation}
f_\nu ^{({\rm III})}=\frac{f^{({\rm I})}}{z-\Omega_{\nu}^2}\ . 
\end{equation}

As well as the poles at $z=0$ and the ($N+1$) roots of the dispersion
relation, this has a pole at $z=\Omega_{\nu}^2$. 
However, the residues at $z=0$ and $z=\Omega_{\nu}^2$
cancel each other, so the sum over the remaining zeros must give zero,
as required.
In summary, we find that
summing over the residues gives:%
\begin{equation}
S_{\nu \nu ^{\prime }}^{({\rm III})}=
\sum {\rm Res}\left[ f_\nu ^{({\rm III})}(z)\right] =0\ . 
\end{equation}

This proves the last sum-rule requirement on the commutators.

\section{Hamiltonian}

We now wish to show that when the Hamiltonian is expressed
in terms of the operators $\hat{a}_{\mu}(k)$ and 
$\hat{a}_{\mu}^{\dagger}(k)$, $\mu=0, \ldots , N,$ it is of
diagonal form.  Our first step is to derive an orthogonality
relation for the classical modes.  This will allow us to show 
that the Hamiltonian contains only terms of the form
$\hat{a}_{\mu}^{\dagger}(k) \hat{a}_{\mu}(k)$.  The next step is
to examine the normalization of the modes.  Once this has been
done we find that the Hamiltonian is given by
\begin{equation}
\label{dham}
H=\sum_{\mu=0}^{N}\int dk\hbar\omega_{\mu}(k)
\hat{a}_{\mu}^{\dagger}(k)\hat{a}_{\mu}(k) .
\end{equation}

In order to find the proper orthogonality relation for the
modes, let us first define the $N+1$ component vector:
\begin{equation}
\overline{\lambda}=\left(\begin{array}{c} \tilde{\lambda} \\
\tilde{p}_{\nu} \end{array}\right),
\end{equation}
or $\lambda_{0}=\tilde{\lambda}$ and $\lambda_{s}=\tilde p_{\nu_{s}}$
for $s \geq 1$.  The eigenvalue equations
can be expressed in the form (for each value of $k$)
\begin{equation}
\label{eigen1}
M\overline{\lambda}=\omega^{2}\overline{\lambda} ,
\end{equation}
where the $(N+1)\times (N+1)$ matrix $M$ is given by
\begin{equation}
M=\left(\begin{array}{cccc} 
k^{2}c^{2} & ikc^{2} & ikc^{2} & \ldots \\
-ikg_{\nu_{1}} & \Omega_{\nu_{1}}^{2} & 0 & \ldots \\
-ikg_{\nu_{2}} & 0 & \Omega_{\nu_{2}}^{2} & \ldots \\
\vdots & \vdots & \vdots &  \\ \end{array}\right) .
\end{equation}
The matrix $M$ is not hermitian, but if it is multiplied 
by the positive, diagonal, $(N+1)\times (N+1)$ matrix $G$,
\begin{equation}
G_{rs}=\delta_{rs}G_{ss} \hspace{1cm} {\rm where} \hspace{5mm}
G_{ss}=\left\{
\begin{array}{c} 1\hspace{5mm} s=0 \\ c^{2}/g_{\nu_{s}}
\hspace{5mm} s\geq 1 \end{array} \right. ,
\end{equation}
then the combination $GM$ is hermitian.  Therefore, if
\begin{equation}
M\overline{\lambda}^{(1)}=\omega_{1}^{2}
\overline{\lambda}^{(1)} \hspace{1cm}M\overline{\lambda}^{(2)}
=\omega_{2}^{2}\overline{\lambda}^{(2)} ,
\end{equation}
then
\begin{eqnarray}
\langle \overline{\lambda}^{(2)}|GM\overline{\lambda}^{(1)}
\rangle & = & \omega_{1}^{2}\langle
\overline{\lambda}^{(2)}|G\overline{\lambda}^{(1)}\rangle \\
 & = & \langle GM\overline{\lambda}^{(2)}|
\overline{\lambda}^{(1)}\rangle =\omega_{2}^{2}\langle
\overline{\lambda}^{(2)}|G\overline{\lambda}^{(1)}\rangle .
\end{eqnarray}
This implies that if $\omega_{1}^{2}\neq \omega_{2}^{2}$, then
\begin{equation}
\langle \overline{\lambda}^{(2)}|G\overline{\lambda}^{(1)}
\rangle = 0 ,
\end{equation}
and we have the desired orthogonality relation.  Expressing
this in slightly more generality, we note that for each value
of $k$ there are $N+1$ eigenvectors $\overline{\lambda}^{(\mu)}$,
$\mu=0, \ldots ,N$, each corresponding to a different 
eigenvalue $\omega_{\mu}(k)$.  As a result we have that
\begin{equation}
\label{orth}
\langle \overline{\lambda}^{(\mu)}|G
\overline{\lambda}^{(\mu^{\prime})}\rangle = \delta_{\mu j^{\prime}}
\langle \overline{\lambda}^{(\mu)}|GM\overline{\lambda}^{(\mu)}
\rangle .
\end{equation}

We now express the fields in terms of the eigenvectors and 
substitute them into the Hamiltonian, which is given by
integrating the Hamiltonian density in Eq. (\ref{honed}) over $x$.  
In particular, we have that 
\begin{eqnarray}
\Lambda_{\mu}=\lambda^{(\mu)}_{0} & \hspace{1cm} & p^{\nu}_{\mu}
=\lambda^{(\mu)}_{\nu} \\
\Pi_{\mu}=-i\mu \omega_{\mu}\lambda^{(\mu)}_{0} & \hspace{1cm} &
\pi^{\nu}_{\mu}=\frac{-i\omega_{\mu}\lambda_{\nu}^{(\mu)}}{\varepsilon_0
g_{\nu}}
 .
\end{eqnarray}
This allows us to use Eqs. (\ref{eigen1}) and (\ref{orth}) when
calculating the Hamiltonian, and we find that 
\begin{equation}
H=4\pi\mu A\sum_{\mu=0}^{N}\int dk \langle
\overline{\lambda}^{(\mu)}(k)|G\overline{\lambda}^{(\mu)}(k)\rangle 
\omega_{\mu}^{2}(k)\hat{a}_{\mu}^{\dagger}(k)\hat{a}_{\mu}(k) .
\end{equation}

In order to show that the Hamiltonian assumes the form given in
Eq. (\ref{dham}), and to justify our assumption that $i
\dot{\hat{a}}_{\mu}=\omega_{\mu}\hat{a}_{\mu}$, we need to prove
that
\begin{equation}
\label{cconda}
4\pi\mu A \langle\overline{\lambda}^{(\mu)}(k)|G
\overline{\lambda}^{(\mu)}(k)\rangle \omega_{\mu}(k) = \hbar .
\end{equation}
Noting that 
\begin{equation}
\lambda_{\nu}^{(\mu)}=\frac{ikg_{\nu}}{\Omega_\nu^{2}-\omega_{\mu}^{2}}
\lambda_{0}^{(\mu)} \hspace{1cm} \lambda_{0}^{(\mu)}=\Lambda_{\mu}
=\sqrt{\frac{\hbar v_{\mu}k}{4\pi\mu A\omega_{\mu}^{2}}} ,
\end{equation} 
we see that Eq.(\ref{cconda}) will be true if 
\begin{equation}
\label{ccond2a}
\left( 1+\sum_{\nu}\frac{k^{2}c^{2}g_{\nu}}{(\Omega_\nu^{2}
-\omega_{\mu}^{2})^{2}}\right)\frac{kv_{\mu}}{\omega_{\mu}} = 1 .
\end{equation}
This implies that
that $v_{\mu}=d\omega_{\mu}/dk$, from
Eq. (\ref{vg1}).  We can then conclude that the expression
for the Hamiltonian given in Eq. (\ref{dham}) is correct.

Summarizing, our theory of a linear medium with $N$ resonances is 
described by the Hamiltonian in Eq. (\ref{dham}) and the 
corresponding field operators have the expansion:%
\begin{equation}
\widehat{D}(t,x)=i\sum\limits_{\mu}\int dk\,k\left[ \frac{\hbar 
\varepsilon(\omega _{\mu}(k))v_{\mu}(k)}{4\pi
k A}\right] ^{1/2}\widehat{a}_{\mu}(k)e^{ikx-i\omega _{\mu}(k)t}+h.c.\
, 
\end{equation}
\begin{equation}
\widehat{E}(t,x)=i\sum\limits_{\mu}\int dk\,\left[ \frac{\hbar 
kv_{\mu}(k)}{4\pi\varepsilon(\omega _{\mu}(k))
A}\right] ^{1/2}\widehat{a}_{\mu}(k)e^{ikx-i\omega _{\mu}(k)t}+h.c.\
, 
\end{equation}
\begin{equation}
\widehat{B}(t,x)=-i\sum\limits_{\mu}\int dk\,\left[ \frac{\hbar\mu 
kv_{\mu}(k)}{4\pi
A}\right] ^{1/2}\widehat{a}_{\mu}(k)e^{ikx-i\omega _{\mu}(k)t}+h.c.\ . 
\end{equation}

Here we have also included the electric field expansion ( which is 
obtained by including the polarization term), for comparison with
more familiar results. As one might expect, the only difference 
between the electric field and displacement field expansions, is
a factor proportional to the dielectric permittivity $\varepsilon(\omega _{\mu}(k)$
in each branch of the dispersion relation. It is important to notice here
that the two fields cannot be related by one, frequency-independent
coefficient. This is a natural consequence of dispersion, and also
occurs in the corresponding classical theory.

In order to provide a more
physical understanding of this result, the summation
over the branches in one dimension can be replaced by an integral
over the mode frequency, in each propagation direction, i.e., define:
 \begin{equation}
\widehat{D}(t,x)= \widehat{D}^{(+)}(t,x)+\widehat{D}^{(-)}(t,x)
\, . 
\end{equation}
Now, since the mode frequency has a
bounded range for each root, we can define a frequency dependent
mode operator as:
\begin{equation}
\widetilde{a}(\omega)=\widehat{a}_{\mu}(k)/\sqrt{|v_{\mu}(\omega)|} \,
, 
\end{equation}
where the appropriate root $\mu$ is chosen in each case to correspond to the
mode frequency argument - except, of course, in the forbidden bands.
The
commutators of the new mode operators are:
\begin{equation}
[\widetilde{a}(\omega),\widetilde{a}^{\dagger}(\omega')] = \delta(\omega
-\omega ')
, 
\end{equation}
and the mode expansion is now the same as it would be in a non-dispersive
case - except that no modes exist in the forbidden bands:
\begin{equation}
\widehat{D}^{(\pm)}(t,x)=\pm i\int_0^{\infty \prime} d\omega\,\left[
\frac{\hbar 
k(\omega)\varepsilon(\omega)}{4\pi
A}\right] ^{1/2}\widetilde{a}(\omega) e^{\pm ik(\omega) x-i\omega t}+h.c.\
\, . 
\end{equation}

Similar equations hold for the other fields; for example, the electric 
and magnetic field
expansions are just:
\begin{equation}
\widehat{E}^{(\pm)}(t,x)=\pm i\int_0^{\infty \prime} d\omega\,\left[
\frac{\hbar 
k(\omega)}{4\pi\varepsilon(\omega)
A}\right] ^{1/2}\widetilde{a}(\omega) e^{\pm ik(\omega) x-i\omega t}+h.c.\
\, . 
\end{equation}

and:
\begin{equation}
\widehat{B}^{(\pm)}(t,x)= - i\int_0^{\infty \prime} d\omega\,\left[
\frac{\hbar \mu 
k(\omega)}{4\pi\varepsilon(\omega)
A}\right] ^{1/2}\widetilde{a}(\omega) e^{\pm ik(\omega) x-i\omega t}+h.c.\
\, . 
\end{equation}

. The important point is that we
can
exactly absorb the group velocity factor into the frequency integral - which,
however,
is only defined in the range of allowed mode frequencies. This 
result (also obtained in earlier narrow-band
Lagrangian approach\cite{CDRS,Drum90}, and in a single-resonance
model\cite{Ho}),
was most clearly emphasized in the single-resonance theory of Huttner and Barnett\cite{Hutt91}.
It
 implies
that the two-time correlation function for narrow-band fields in the
transmission
band, is essentially identical to those of the corresponding vacuum fields,
apart from the usual reflectivity factors. This is a necessary ingredient of
any
theory of the interface properties of the fields, and will be explored
in more detail in a subsequent paper. The above mode expansion has been
widely used in quantum optics, and the present result shows that it is
exact for a realistic, multiple-resonance model of a dispersive medium - provided we recognize
that there are no modes in the forbidden bands.

\section{Higher-dimensional models}

We next consider an
n-dimensional Lagrangian for oscillators with mass $m_\nu$, displacement ${\bf
r}_\nu$,
effective charge $q_\nu$, 
and  oscillation frequencies $\Omega_\nu $
about their  center of charge
 position $\overline{\bf x}_\nu $:
\begin{equation}
{\cal L}_a=\sum\limits_\nu \left[ \frac {{m}_{\nu}} {2}\left( \dot {{\bf r} }%
_{\nu}^2-\Omega^2_{\nu}{\bf r} _{\nu}^2\right) +\frac {q_\nu}{\varepsilon _0}
{\bf r} _\nu \cdot {\bf D}(%
\overline{\bf x}_\nu )\right] \ . 
\end{equation}
Here $[{\bf r} _\nu q_\nu] $ is the dipole moment of a
charged particle  in a multi-polar
Drude-Lorentz model. In this general case, the
labels ${\nu}$ may correspond either to  distinct resonances
of one atom
or  to  distinct positions. Each resonance has
its own corresponding effective charge -
and hence dipole moment. Any sum rules are incorporated into
the definitions of the effective charges involved in a given
transition. For
simplicity, the self-energy terms proportional to ${\bf P}^2$ are included
in the definition of the resonant frequencies,
which are defined to diagonalize the individual charge-cell Hamiltonians
in the multipolar gauge.
The coupling in multipolar gauge is
to the displacement field ${\bf D}(\overline{\bf x}_\nu )$ at the central
position
$\overline{\bf x}_\nu$,
used as an origin for defining a local polarization.
All inter-atomic Coulomb terms in this gauge, are carried by the
photon-exchange process,
which couples atoms at distinct spatial positions.

We introduce a vector gauge field - the dual potential 
${\bbox \Lambda} $%
, so that ${\bf D}=\nabla \times {\bbox \Lambda}$ and 
${\bf B}=\mu  \partial 
{\bbox \Lambda}/ \partial t$. 
The discrete cell positions are now replaced by a continuum 
polarization field as before, together with the appropriate 
 local field corrections. 
To account for more general dielectric structures that
may have local interactions not included in the Coulomb corrections,
we now include a quadratic dispersion term $\alpha_{\nu}$
to describe
the residual phonon and exciton dispersion\cite{Phdisp},
that exists
in the absence of long wavelength electromagnetic couplings.

The simplest Lagrangian density that generates
the correct electromagnetic energy and Maxwell's equations 
for an $n$-dimensional wave-guide with effective
cross-section $A \simeq d^{3-n}$ is: \begin{eqnarray}
{\cal L}/A&=&{1 \over 2\mu}{\dot 
{\mathbf \Lambda} }^2({\bf x})+
{1 \over \varepsilon _0}\left[
{\bf P} ({\bf x}) \cdot
\nabla \times {\mathbf \Lambda} ({\bf x})
-
{1 \over 2}(\nabla \times
{\mathbf \Lambda}({\bf x}) )^2 \right] \\ \nonumber
&+& 
\sum\limits_{\nu=1}^N \left[ 
\dot {\bf p}_{\nu}^2({\bf x})
-\Omega^2_{\nu}{\bf p}_{\nu}^2({\bf x})
-\alpha_{\nu}({\bf x})[\nabla_i{\bf p}_{\nu}({\bf x})]^2
\right]/({2 \varepsilon_0 g_\nu({\bf x})}) \,
. 
\end{eqnarray}

Here the 
polarization density due to all the  dipoles
is  ${\bf P}({\bf x})=\sum_\nu {\bf p}_\nu({\bf x})=
\sum_\nu {\bf r}_\nu q_\nu
\rho _\nu ({\bf x})$, where
$\rho _\nu ({\bf x})$ is the number 
density of the oscillators with resonant
frequency $\Omega_{\nu}  $.  The partial 
polarization ${\bf p} _\nu({\bf x}) $ is  regarded as a continuum field, with
$\nu = 1,.. N$ labeling the bare frequency
of the elementary phonon and exciton resonances, in the absence
of coupling to the long wavelength photons.
We have assumed that the dispersion
of phonon and exciton modes are the same for longitudinal and transverse
modes. The transverse dispersion will ultimately be modified by the
coupling between the field and the polarization, which is entirely
included in the term $g_\nu ({\bf x})$, as defined in Eq.
(\ref{gnu}).

We impose the usual gauge constraint that $\nabla \cdot{\bbox \Lambda}=0$,
so that the field variable only has transverse degrees of freedom;
this does not apply to the polarization. In addition,
we can impose
wave-guiding conditions that ${\bbox \Lambda}$ is restricted to
a one, two or three dimensional manifold. In practise,
dispersion occurs in the electromagnetic mode functions 
(which are wavelength dependent), so that
it is necessary to solve for the complete three-dimensional
mode structure to rigorously treat a fiber waveguide, for example. However,
a simple low-dimensional wave-guiding theory is still useful
 as a guide to the behavior of a complete theory. 
 In the full three-dimensional case, the area term $A$ is simply omitted, as $A=1$.

The resulting generalized Maxwell-Bloch equations are:%
\begin{eqnarray}
\left[{\partial_ t^2}-c^2\nabla^2\right]{\bbox \Lambda} 
&=&c^2\nabla \times {\bf P}({\bf x}) \nonumber \\
\left[{\partial_t^2}+ \Omega^2_{\nu}(k)\right]{\bf p}_\nu 
&=&g_\nu({\bf x}) \nabla \times{\bbox \Lambda}
+ \nabla_i \left(\alpha_{\nu}({\bf x})
 \nabla_i{\bf p}_\nu({\bf x})\right)
 \ . 
\end{eqnarray}

In order to quantize the system, we  introduce the
canonical momenta , ${\bbox \Pi }({\bf x})$ and 
${\bbox \pi} _\nu({\bf x})$, where:%
\begin{eqnarray}
{\bbox \Pi }({\bf x}) =\mu \dot{{\bbox \Lambda} }({\bf x}) ={\bf B}({\bf x})\ , 
\nonumber \\
{\bbox \pi} _\nu({\bf x}) ={ 1 \over \varepsilon _0g_\nu({\bf x})}
 \dot{{\bf p} }_\nu ({\bf x}) \propto m{\bf v}\ . 
\end{eqnarray}

The quantization rules are the usual ones
obtained on replacing Poisson brackets with operator commutators.
Scaling by $A$ is introduced so that the field units are 
independent of waveguide dimension, and the delta-functions are
n-dimensional.
All position-type operators
of the same type of variable must commute amongst themselves,
as do all momentum-type operators. The commutators
involving cross-terms between the position and momentum operators,
and between the different types of variable, are:

\begin{eqnarray}
({\rm I})&\ \ \ \ \ \ 
\left[ \widehat{{D} }_i({\bf x})\,,\,\widehat{ B }_{j}({\bf x}^{\prime
})
\right] &=i\hbar \nabla_x \times
\delta^{\perp}_{i j}({\bf x }-{\bf x}^{\prime })/A\ , 
\nonumber  \\
({\rm II})&\ \ \ \ \ \
\left[ \widehat{{ p} }_{i\nu } ({\bf x})\,,\,\widehat{ \pi }_{j \nu ^{\prime
}}({\bf x}^{\prime })\right] &=i\hbar\delta_{i j} \delta_{\nu \nu ^{\prime }}
\delta({\bf x}-{\bf x}^{\prime })/A\ , 
\nonumber  \\
({\rm III})&\ \ \ \ \ \
\left[ \widehat{{D} }_i({\bf x})\,,\,
\widehat{{ p} }_{j \nu}({\bf x}^{\prime })\right] &=0 \ , 
\nonumber  \\
({\rm IV})&\ \ \ \ \ \
\left[ \widehat{ B }_i({\bf x})\,,\,\widehat{ \pi }_{j \nu }({\bf x}^{\prime
})\right]
 &=0 , 
\nonumber  \\
({\rm V})&\ \ \ \ \ \
\left[ \widehat{{ D} }_i({\bf x})\,,\,
\widehat{ \pi }_{j \nu }({\bf x}^{\prime })\right] &=0 \ , 
\nonumber  \\
({\rm VI})&\ \ \ \ \ \
\left[ \widehat{ B }_{i  }({\bf x})\,,\,
\widehat{{ p} }_{j \nu }({\bf x}^{\prime })\right] &=0 \ . 
\end{eqnarray}

Here we have introduced the usual notation of a transverse delta function
defined so that:
\begin{equation}
\delta^{\perp}_{i j}({\bf x}) = {1 \over (2 \pi )^n}\int d^n{\bf k}
\tilde\delta^{\perp}_{i j}({\bf k})e^{i{\bf k} \cdot {\bf x}} \ ,
\end{equation}
where 
$\tilde\delta^{\perp}_{i j}({\bf k}) \equiv (\delta_{i j}-k_i k_{j}/|{\bf
k}|^2)$
is the transverse projector in momentum space. In three dimensions,
the first commutator can also
be written in the more familiar form of:
\begin{equation}
({\rm I})\ \ \ \ \ \ 
\left[ \widehat{{E} }_i({\bf x})\,,\,\widehat{ B }_{j}({\bf x}^{\prime
})
\right] =i\hbar \nabla_x \times
\delta^{\perp}_{i j}({\bf x }-{\bf x}^{\prime })/\varepsilon_0 \ . 
\end{equation}

The final quantum Hamiltonian density is (using normal ordering):
 \begin{eqnarray}\label{Hdensity}
{\cal H}/A&=&
 {1\over 2\varepsilon _0} \widehat{{\bf D} }^2({\bf x})  
+{1\over 2 \mu } \widehat{{\bf B} }^2({\bf x})  
-
{1\over \varepsilon _0}\widehat{{\bf D} }({\bf x}) \cdot \widehat{{\bf P} }({\bf x}) 
 \nonumber \\ &+& 
{1\over 2\varepsilon _0 g_\nu({\bf x})   }\sum\limits_{\nu =1}^N 
\left[ \varepsilon^2 _0 g_\nu^2 ({\bf x})  
\widehat{\mathbf \pi }_{\nu}^2({\bf x}) +
\Omega^2_{\nu}\widehat{{\bf p} }_\nu^2 ({\bf x})
+\alpha_{\nu}({\bf x})  [\nabla_i{\bf p}_{\nu}({\bf x})]^2
 \right]  
\end{eqnarray}

These equations  hold for an arbitrary spatial distribution 
$\rho({\bf x})$ of
the continuum of polarizable atoms. 

\section{Three-dimensional Quantum mode operators}

In this section, we specialize to the case of a
continuum model with uniform couplings and velocities, 
as in the one-dimensional case,
in order to
find the dispersion relations for plane-waves.
As in the one-dimensional case as well,
we introduce a mode structure by finding the eigenmodes 
 to Maxwell's equations in the form:%
\begin{equation}
\left( 
\begin{array}{c}
{\bbox \Lambda} (t,{\bf x}) \\ 
{\bf p}_\nu (t,{\bf x})
\end{array}
\right) =\left( 
\begin{array}{c}
\widetilde{{\bbox \Lambda} }(\omega ,{\bf k})\\ \widetilde{\bf p}_\nu (\omega
,{\bf k})
\end{array}
\right) e^{i{\bf k\cdot x}-i\omega t}\ . 
\end{equation}

Defining $k=|{\bf k}|$, 
it follows that these satisfy the Maxwell-Bloch equations in the form:%
\begin{eqnarray} \label{mode3d}
\left( -\omega ^2+c^2k^2\right) \widetilde{{\bbox \Lambda} }=ic^2{\bf k}\times
\sum\limits_{\nu}\widetilde{\bf p}_{\nu}\ , 
\nonumber \\
(-\omega ^2+\Omega^2_{\nu}(k))\widetilde{\bf p}_\nu =ig_\nu {\bf k}\times
\widetilde{{\bbox \Lambda} }\ . 
\end{eqnarray}

Our model now includes simple
phonon/exciton dispersion. These effects cause the Fourier-domain
equations for the polarization to be modified, so that
$\Omega^2_{\nu}$ is now replaced by a momentum
dependent function $\Omega^2_{\nu}(k) = \Omega^2_{\nu}+k^2 
\alpha_{\nu}$, where we define $k=|{\bf k}|$ in this section.
The phonon/exciton  dispersion can be generalized to more complicated cases if desired,
with higher order k-dependences -
although in many cases only the relatively long wavelength (small $k$)
phonons are relevant to optical properties. 
In terms of the coupled equations given above, only
the transverse part of the polarization couples to the photons,
hence we can define:
\begin{equation}
\widetilde{ p}_{i}^{\perp}({\bf k})  =  \tilde\delta^{\perp}_{i j}({\bf k})
\widetilde{ p}_{j}({\bf k})  \ .
\end{equation}

Solving for 
$\widetilde{\bf P}^{\perp}$ by summing over the 
transverse polarizations of all the oscillators,
we find:
\begin{equation}
(c^2k^2-\omega ^2)\widetilde {\bf P}^{\perp}=\widetilde {\bf P}^{\perp}
c^2k^2\sum\limits_\nu \frac{g_\nu }{\Omega^2_{\nu}(k)-\omega ^2}\
. 
\end{equation}

The eigenvalues of the longitudinal part of the polarization
field are not changed by the coupling, while
eliminating the 
transverse polarization field $\widetilde {\bf P}^{\perp}$ 
leads to the eigenvalue equation:%
\begin{equation}
{\omega ^2} =\frac{c^2k^2}
{n^2(\omega )}\equiv c^2k^2
\left[1-\sum\limits_\nu \frac{g_\nu }{\Omega^2_{\nu}(k)-\omega ^2}\right]
\, . 
\end{equation}

For any wave-number $k=|{\bf k}|$, we find a band-structure in which there are
typically ($2(N+1)$) positive roots $\omega_{0\sigma}(k),..
\omega_{N\sigma}(k)$
to the transverse eigenvalue equations
for each ${\bf k}$ value, with $\sigma = 1,2$, and $\omega_{\mu 1}(k)
=\omega_{\mu 2}(k)$
due to the isotropy of our model. Each mode therefore has
 two orthogonal polarization unit vectors ${\bf u}_{\sigma}({\bf k})$,
 such that ${\bf k}\cdot{\bf u}_{\sigma}({\bf k}) = 0$.
Similarly, there are ($N$) longitudinal roots, which are labeled $\sigma = 0$,
and are unchanged by the long wavelength electromagnetic couplings.
In the case where phonon/exciton dispersion is omitted, the transverse
dispersion relation is precisely the same as in the one-dimensional model.

As before, the solution is unique for any given modal frequency, but has
forbidden
regions which
indicate a resonance, or absorption band. 
Typical dispersion relations for this model
also
demonstrate the existence of transmission and absorption
bands - but with additional structure in each branch, whose origin is
in the phonon (or exciton) dispersion. From now on,
 we use the notation $\omega_{\mu\sigma}(k)$ with $\sigma =1,2$,
 to indicate a solution to the
 full transverse equations. The notation 
 $\omega_{\mu0}(k)=\Omega_{\mu}(k)$ indicates the longitudinal eigenvalue,
 which of course is undefined for $\mu = 0$, in the absence of longitudinal
photons.

\subsection{Three-dimensional commutators}

Having derived the modal solutions, we now wish to expand the fields of the
theory in terms of annihilation and creation operators in the
three-dimensional
model. 
These have the function of diagonalizing the Hamiltonian, which
we anticipate
will have the final structure of:
\begin{equation}
H=\sum\limits_{\mu=0}^{N}\sum\limits_{\sigma=0'}^{2}\int d^n{\bf k}\hbar 
\omega _{\mu\sigma}({k})\widehat{a}_{\mu\sigma}^{\dagger }({\bf k})%
\widehat{a}_{\mu\sigma}({\bf k})\, .
\label{H3D} 
\end{equation}
Here the lower limit notation of $\sigma=0'$ is defined to exclude the
unphysical
combination of $\mu=0,\sigma =0$.
As before, the modal frequency
$\omega _{\mu\sigma}({k})$ is the inverse of $k(\Omega )$, for the
$\mu$-th transverse branch.
This expansion requires  that we
define mode operators $\widehat{a}_{\mu\sigma}$ in the $\mu$-th branch
of the dispersion relation so that:%
\begin{equation}
\widehat{{\bbox \Lambda} }(t,{\bf x})=
\sum\limits_{\mu=0}^{N}
\sum\limits_{\sigma=1}^{2}\int d^n{\bf k}\left[ {\bbox \Lambda}
_{\mu\sigma}({\bf k})%
\widehat{a}_{\mu\sigma}({\bf k})e^{i{\bf k\cdot x}-i\omega
_{\mu\sigma}({k})t}+h.c.\right] \ . 
\end{equation}
The summation here is just over the transverse
 branches in the dispersion relation. 
These combined transverse field-polarization excitations can be termed
polaritons, and we notice here that there can be longitudinal excitations
that propagate, as is usual in solid-state physics.
The commutation properties of the annihilation and creation operators
$\widehat{a}_{\mu\sigma}({\bf k})$ are chosen so that:%
\begin{equation}
\left[ \widehat{a}_{\mu\sigma}({\bf k}),
\widehat{a}_{\mu'\sigma'}^{\dagger }({\bf k}^{\prime })\right]
=\delta_{\mu\mu'}
\delta_{\sigma \sigma'}\delta({\bf k}-{\bf k}^{\prime })\ . 
\end{equation}

Similarly, the momentum field can be expanded as:%
\begin{equation}
\widehat{\bbox \Pi }(t,{\bf x})=
\sum\limits_{\mu=0}^{N}
\sum\limits_{\sigma=1}^{2}\int d^n{\bf k}
\left[ {\bbox \Pi }_{\mu\sigma}({\bf k})\widehat{a}%
_{\mu\sigma}({\bf k})e^{i{\bf k\cdot x}-i\omega
_{\mu\sigma}({k})t}+h.c.\right] \ . 
\end{equation}

Thus, at equal
times:%
\begin{equation}
\left[ \widehat{\Lambda}_i({\bf x})\,,\,
\widehat{\pi }_{j}({\bf x}^{\prime })\right] =i\hbar
\delta^{\perp}_{i j}({\bf x}-{\bf x}^{\prime })/A\ =
\sum\limits_{\mu=0}^{N}
\sum\limits_{\sigma=1}^{2}
\int d^n{\bf k}\left[ { \Lambda}
_{i\mu\sigma}({\bf k})\Pi _{j\mu\sigma}^{*}({\bf k})
e^{i{\bf k}\cdot({\bf x}-{\bf x}^{\prime })}-h.c.\right] \ . 
\end{equation}

This implies  that, in order to preserve commutation
relations, we have similar results to the one-dimensional case -%
\begin{equation}
\sum\limits_{\mu=0}^{N}
\sum\limits_{\sigma=1}^{2}{\Lambda} _{i\mu\sigma}({\bf k})
\Pi _{j\mu\sigma}^{*}({\bf k})=\frac{i\hbar }{2A(2\pi)^n}
\tilde\delta^{\perp}_{i j}({\bf k})\ . 
\end{equation}

For free fields, it is clear that ${\bbox \Pi} =\mu\dot 
{\bbox \Lambda} $. Hence, we can also write ${\bbox \Pi} _{\mu\sigma}({\bf
k})$
in the form of:%
\begin{equation}
{\bbox \Pi} _{\mu\sigma}({\bf k})=-i\omega _{\mu\sigma}({k})\mu {\bbox
\Lambda} _{\mu\sigma}({\bf k})\ . 
\end{equation}

The equation for the expansion coefficients ${\bbox \Lambda} _{\mu\sigma}({\bf
k})$,
is therefore:%
\begin{equation}
\sum\limits_{\mu=0}^{N}\sum\limits_{\sigma=1}^{2}
\omega _{\mu\sigma}({k})
{\Lambda} _{i\mu\sigma}({\bf k})
{\Lambda} _{j\mu\sigma}({\bf k})=
\frac \hbar {2 A\mu(2\pi)^n }\tilde\delta^{\perp}_{i j}({\bf k})\ .  
\end{equation}

Next, we  choose ${\bbox \Lambda}
_{\mu\sigma}({\bf k})$ to be real, and
as before, we can always choose an
(unknown) function $v_{\mu\sigma}(k)$ so that:
\begin{equation}
{\bbox \Lambda} _{\mu\sigma}({\bf k})=
{\bf u}_{\sigma}({\bf k})
\left[ \frac{\hbar v_{\mu\sigma}(k)\varepsilon_{\mu}({k})}{2Ak
(2\pi)^n}\right]
^{1/2}\ .
\label{Lambdamus} 
\end{equation}

We will show later that $v_{\mu\sigma}(k)$ must be interpreted
as
the electromagnetic component of group velocity,
with non-electromagnetic 
phonon/exciton dispersion explicitly excluded. This is not completely identical
to
either
the
 earlier narrow-band analysis\cite{Drum90}
of this problem, 
or the simple one-dimensional theory.
The difference can be attributed to the difference
in the Hamiltonian energy that is introduced when the polarization
fields are dispersive.

In order to demonstrate this, we first 
recall the standard identity that:
\begin{equation}
\sum\limits_{\sigma=1}^{2}
{ u}_{i\sigma}({\bf k}){ u}_{j\sigma}({\bf k})
=
\tilde\delta^{\perp}_{i j}({\bf k})\ .  
\end{equation}
Next, substituting the above expression
for ${\bbox \Lambda} _{\mu\sigma}({\bf k})$ into the equation for
 the field and
mode commutators leads to:%
\begin{equation}
({\rm I})\ \ \ \ \ \ \ \ \ \ \ \
\sum\limits_{\mu=0}^{N}\frac{kv_{\mu\sigma}(k)}{%
\omega _{\mu\sigma}({k})}=1\, . 
\end{equation}

As explained above, 
we have to determine a function $v_{\mu\sigma}(k)$ which satisfies this
condition,
and we intend to demonstrate that the choice of $v_{\mu\sigma}(k)$
as the (isotropic) electromagnetic component of group-velocity of the relevant
polariton
branch is sufficient --
no other correction factors are needed. At this point, we notice an important
fact; apart from the change in the resonance frequencies $\Omega _{\nu}({k})$
due to their k-dependence, the above summation over
$\omega _{\mu\sigma}({k})$ is identical in analytic form to our
one-dimensional expression. However, the k-dependence plays no role
in obtaining the Cauchy theorem summation results, provided we define
$v_{\mu\sigma}(k)$
to be the same function of $\Omega _{\nu}({k})$ and $\omega _{\mu\sigma}({k})$ 
as before. In other words, the
 k-dependence of the phonon-exciton dispersion relation
 simply renormalizes the effective resonance frequency at each k-value, in
  the above summation over the roots of the dispersion relation. Since this
  renormalization is the same for each eigenvalue, the summation
  can be carried out using identical techniques to those used previously.
  Thus, we define, for $\sigma = 1,2$:
   \begin{equation}
\label{vg2}
v_{\mu\sigma}(k) = \frac{\Omega_{\mu}(k)}{k}
\left( 1+\sum_{\nu}\frac{k^{2}c^{2}g_{\nu}}
{(\Omega_{\nu}^{2}(k)
-\omega_{\mu\sigma}(k)^{2})^{2}}\right)^{-1} .
\end{equation}
It should be noted that this $v_{\mu\sigma}(k)$
is {\em not} the same as $v_{\mu\sigma}(k)$ in the one-dimensional
case, although it has the same functional form. The difference is that it is
a now a function of $\Omega_{\mu}(k)$ and $\omega_{\mu\sigma}(k)$, which
include
phonon/exciton  dispersion effects. However, if we simply differentiate the slope of
the function $\omega_{\mu\sigma}(k)$, we do not obtain $v_{\mu\sigma}$ as
defined
here - there are additional terms involving
$\partial\Omega_{\nu}^{2}(k)/\partial k$.
For this reason, we refer to $v_{\mu\sigma}(k)$ as the electromagnetic
component
of group velocity, which excludes additional  transport terms.

It is clear that the mode function expansion of $%
\widehat{{\bf p} }_\nu ,\,\widehat{\bbox \pi }_\nu $ is also needed,
for a complete demonstration of consistency. Using
Maxwell's equations,  if we expand $\widehat{{\bf p} }_\nu $ as:%
\begin{equation}
\widehat{{\bf p} }_\nu =\sum\limits_{\mu=0}^{N}
\sum\limits_{\sigma=0}^{2}\int d^n{\bf k} \left[ {\bf p} _{\mu\sigma}^\nu
({\bf
k})%
\widehat{a}_{\mu\sigma}({\bf k})e^{i{\bf k}\cdot{\bf  x}-i\omega
_{\mu\sigma}({k})t}+h.c.\right] \ , 
\end{equation}
then it follows that the expansion coefficient for 
the transverse component of ${\bf p} _{\mu\sigma}$
in the $\mu$-th frequency band
must be:
\begin{equation}
{\bf p}_{\mu\sigma}^\nu ({\bf k})=\frac{ig_{\nu}{\bf k}\times{\bbox \Lambda}
_{\mu\sigma}({\bf k})}
{(\Omega^2_{\nu}(k)-\omega^2_{\mu\sigma}(k))}%
\ . 
\label{pmus}
\end{equation}

Similarly, if the canonical momentum for the atomic polarization field is:%
\begin{equation}
\widehat{\bbox \pi }_\nu (t,{\bf x})=\sum\limits_{\mu=0}^{N}
\sum\limits_{\sigma=0}^{2}\int d^n{\bf k}\left[ \pi _{\mu\sigma}^\nu ({\bf
k})%
\widehat{a}_{\mu\sigma}({\bf k})e^{i{\bf k\cdot x}-i\omega
_{\mu\sigma}({k})t}+h.c.\right] \ , 
\end{equation}
then the corresponding transverse expansion coefficient is:%
\begin{equation}
{\bbox \pi} _{\mu\sigma}^\nu ({\bf k})=\frac{ \omega _{\mu\sigma}({k})
{\bf k} \times{\bbox \Lambda} _{\mu\sigma}({\bf k})}{\varepsilon
_0(\Omega_\nu^2(k)
-\omega _{\mu\sigma}^2(k))}\ . 
\end{equation}

For these operators to have the correct equal-time commutators, 
 the different oscillator position operators
 $\widehat{{\bf p} }_\nu$ must commute amongst themselves
 at equal times, as must the
 different momentum operators $\widehat{\bbox \pi }_\nu $. This is trivial
 from the form of the 
 mode operator expansion. However, the commutation
 relations (II) between the position and momentum operators are non-trivial.
The relevant commutation
conditions are therefore:%
\begin{equation}
\left[ \widehat{{ p} }_{i\nu} ({\bf x})\,,\,\widehat{ \pi }_{j\nu ^{\prime
}}({\bf x}^{\prime })\right] 
=i\hbar \delta_{\nu \nu ^{\prime }}\delta_{i j}
\delta({\bf x}-{\bf x}^{\prime })/A\ 
=
\sum\limits_{\mu=0}^{N}
\sum\limits_{\sigma=0'}^{2}
\int d^n{\bf k}\left[ { p} _{i\mu\sigma}^\nu
({\bf k})\pi _{j\mu\sigma}^{*\nu ^{\prime }}({\bf k})
e^{i{\bf k}\cdot({\bf x}-{\bf x}^{\prime })}-h.c.\right] \ . 
\end{equation}
This in turn implies that:%
\begin{equation}
\sum\limits_{\mu=0}^{N}
\sum\limits_{\sigma=0'}^{2}
{ p} _{i\mu\sigma}^\nu ({\bf k})\pi _{j\mu\sigma}^{*\nu ^{\prime }}({\bf k})
=\frac{%
i\hbar \delta_{\nu \nu ^{\prime }}\delta_{i j}}{2A(2\pi)^n}\ . 
\end{equation}

Defining $\delta^{\parallel}_{i j}({\bf k})=
k_ik_j/k^2 = \delta_{i j}-\delta^{\perp}_{i j}({\bf k})$,
and
expanding the coefficients, gives two new equations. The transverse case is:%
\begin{equation}
({\rm II})\ \ \ \ \ \ \ \
\sum\limits_{\mu=0}^{N}\frac{c^2k^3v_{\mu\sigma}(k)g_{\nu}}
{\omega _{\mu\sigma}({k})(\omega
_{\mu\sigma}^2(k)-\Omega^2_{\nu}(k))
(\Omega_{\mu\sigma}^2({\bf k})-\Omega^2_{\nu ^{\prime }})}=\delta_{\nu \nu
^{\prime }}\ , 
\end{equation}
while the longitudinal equation is easily solved
on defining ${\bf u}_0({\bf k}) = {\bf k}/k$: 
\begin{equation} \label{pmulong}
 {\bf p} _{\mu 0}^\nu ({\bf k}) = {\bf u}_0({\bf k})\delta_{\mu \nu}
\bigg[{\hbar g_{\mu} \epsilon_0 \over  2A(2\pi)^n \Omega_{\mu}(k)}\bigg]^{1/2}
\ . 
\end{equation}

Finally, to ensure that 
there are correct field-atom commutators, we must 
satisfy the commutators III-VI. For these cross-terms between the
oscillators and field variables, we find that conditions (III)
and (IV), involving commutators between the field and the 
particle position (or the field momentum and particle momentum)
are automatically satisfied. 
This occurs for the same reason that commutators like 
$[\widehat{{\bbox \Lambda} }({\bf x}),\widehat{{\bbox \Lambda} }({\bf x}')]$ 
or $[\widehat{\bbox \pi }({\bf x}),\widehat{\bbox \pi }({\bf x}')]$ must equal
zero
in our mode expansion. In all these cases involving pairs of 
canonical position-type operators or pairs of momentum-type
operators, the commutator reduces to an odd integral over $k$,
which  vanishes when integrated over
all positive and negative k-values.

This leaves the 
requirements (V) and (VI), which are that
$\widehat{{\bbox \Lambda} }$ and $\widehat{\bbox \pi }_\nu$ must commute at
equal times ,
as well as $\widehat{\bbox \pi }$ and $\widehat{{\bf r}}_\nu$. These
two requirements {\it both} imply the same restriction
on the expansion coefficients, and hence on $v_{\mu\sigma}(k)$,
which is that
 for all $k$ and $\nu $ we must have the condition:%
\begin{equation}
({\rm V},{\rm VI})\ \ \ \ \ \ \ \ \ \ \ \ \sum\limits_{\mu=0}^{N}\frac{
kv_{\mu\sigma}(k)}{\omega _{\mu\sigma}^2(k)
(\omega _{\mu\sigma}^2(k)-\Omega^2_{\nu}(k))}=0\ .
 \end{equation}

Despite the complex nature of each of these conditions - which involve
sums over all the roots of the dispersion equation, and must be satisfied
for all the resonant frequencies $\omega_{\mu\sigma}$, as well as all momenta
$k$ -
we will show that each of these sums can be analytically evaluated
without requiring an algebraic solution for the roots, just as before. 
For all of the 
commutation relation identities it is preferable to use techniques from
complex
function theory,
which transform the sums over roots of the dispersion relation
to complex contour integrals of related meromorphic functions.
However, the 
transverse dispersion relations considered here have an
identical analytic structure for a fixed k-value, with those in
in the one-dimensional case, so the previous analytic results follow without
any further calculation. The main point here is that it is necessary
for the `group-velocity' coefficient to have the  algebraic form
 given in Eq.(\ref{vg2}) - which implies that it includes only part
 of the slope of the dispersion relation.

\section{Hamiltonian}

We now wish to show that when the Hamiltonian is expressed
in terms of the operators $\hat{a}_{\mu\sigma}({\bf k})$ 
and $\hat{a}_{\mu\sigma}^{\dagger}({\bf k})$, $\mu =0, 
\ldots , N$ and $\sigma = 0,1,2$, where the combination
$\mu= \sigma = 0$ is omitted, it is of diagonal form.  
In order to prove this we shall study further the classical
modes, in particular their orthogonality and normalization
properties.  Once this has been done, we shall be able to 
show that the Hamiltonian takes the form given in 
Eq. (\ref{H3D}).

Let us begin by restating the classical mode equations,
Eqs. (10), in matrix form.  Define the $3N+3$ component
vector $\overline{{\bf \lambda}}$ by 
\begin{equation}
\overline{{\bf \lambda}}=\left(\begin{array}{c} 
\tilde{{\bf \Lambda}} \\
\tilde{{\bf p}}_{\nu_{1}} \\ \vdots \\ 
\tilde{{\bf p}}_{\nu_{N}}
\end{array}\right),
\end{equation}
or ${\bf \lambda}_{0}=\tilde{{\bf \Lambda}}$ and 
${\bf \lambda}_{s}={\bf p}_{\nu_{s}}$
for $s \geq 1$, and the $3(N+1)\times 3(N+1)$ matrix 
$M({\bf k})$ by
\begin{equation}
M({\bf k})=\left(\begin{array}{cccc} 
k^{2}c^{2}I_{3} & -ic^{2}K & -ic^{2}K & \ldots \\
-ig_{\nu_{1}}K & \Omega^{2}_{\nu_{1}}(k)I_{3} & 0 & \ldots \\
-ig_{\nu_{2}}K & 0 & \Omega^{2}_{\nu_{2}}(k)I_{3} & \ldots \\
\vdots & \vdots & \vdots &  \end{array}\right) .
\end{equation}  
Here $I_{3}$ is the $3\times 3$ identity matrix and 
$K({\bf k})$ is
the anti-hermitian matrix given by
\begin{equation}
K = \left(\begin{array}{ccc}
0 & -k_{3} & k_{2} \\
k_{3} & 0 & -k_{1} \\
-k_{2} & k_{1} & 0 \end{array} \right),
\end{equation}
which has the action on an arbitrary vector ${\bf A}$
\begin{equation}
K{\bf A}={\bf k}\times {\bf A} .
\end{equation}
The equations for the modes can now be expressed as
(for each value of ${\bf k}$)
\begin{equation}
\label{eigenx}
M\overline{{\bf \lambda}}=\omega^{2}\overline
{{\bf \lambda}} .
\end{equation}

The matrix $M$ is not hermitian, but if it is multiplied 
by the positive, diagonal, $3(N+1)\times 3(N+1)$ matrix $G$,
\begin{equation}
G =\left(\begin{array}{cccc} 
I_{3} & 0 & 0 & \ldots \\
0 & \frac{c^{2}}{g_{\nu_{1}}}I_{3} & 0 & \ldots \\
0 & 0 & \frac{c^{2}}{g_{\nu_{2}}}I_{3} & \ldots \\
\vdots & \vdots & \vdots &  \end{array} \right) ,
\end{equation}
then the combination $GM$ is hermitian.  Therefore, if
\begin{equation}
M\overline{{\bf \lambda}}^{(1)}=\omega_{1}^{2}
\overline{{\bf \lambda}}^{(1)} \hspace{1cm}M
\overline{{\bf \lambda}}^{(2)}
=\omega_{2}^{2}\overline{{\bf \lambda}}^{(2)} ,
\end{equation}
then
\begin{eqnarray}
\langle \overline{{\bf \lambda}}^{(2)}|GM
\overline{{\bf \lambda}}^{(1)}
\rangle & = & \omega_{1}^{2}\langle \overline{{\bf
\lambda}}^{(2)}|G\overline{{\bf \lambda}}^{(1)}\rangle \\
 & = & \langle GM\overline{{\bf \lambda}}^{(2)}|
\overline{{\bf \lambda}}^{(1)}\rangle =\omega_{2}^{2}
\langle \overline{{\bf \lambda}}^{(2)}|G
\overline{{\bf \lambda}}^{(1)}\rangle .
\end{eqnarray}
This implies that if $\omega_{1}^{2}\neq \omega_{2}^{2}$, 
then
\begin{equation}
\label{orth1}
\langle \overline{{\bf \lambda}}^{(2)}|G
\overline{{\bf \lambda}}^{(1)} \rangle = 0 ,
\end{equation}
and we have part of the desired orthogonality relation.  

In order to learn more we must examine the $3N+3$ 
eigenvectors in more detail.  Define the projection
operator,  which projects
each component of $\overline{{\bf \lambda}}$ onto its
longitudinal component,
\begin{equation}
P(\hat{{\bf k}})=\left( \begin{array}{ccc}
|\hat{{\bf k}}\rangle\langle \hat{{\bf k}}| & 0 & 
\ldots \\
0 & |\hat{{\bf k}}\rangle\langle \hat{{\bf k}}| & 
\ldots \\
\vdots & \vdots &  \end{array} \right) ,
\end{equation}
where $|\hat{{\bf k}}\rangle\langle \hat{{\bf k}}|$ is the
projection onto the vector $\hat{{\bf k}}$.  
A short calculation shows that 
$[P(\hat{{\bf k}}),M]=0$ which implies that the eigenvectors
$\overline{{\bf \lambda}}$ can be taken to lie in either 
the subspace projected out by $P(\hat{{\bf k}})$ 
(longitudinal modes), or in the orthogonal subspace 
(transverse modes).  The longitudinal modes can be
found by taking the inner product of Eqs. (\ref{mode3d}) with
$\hat{{\bf k}}$ giving
\begin{equation}
(k^{2}c^{2}-\omega^{2})\hat{{\bf k}}\cdot\tilde{{\bf \Lambda}}
=0 \hspace{1.5cm} (\Omega_{\nu}^{2}(k)-\omega^{2})
\hat{{\bf k}}\cdot\tilde{{\bf p}}_{\nu}=0 .
\end{equation}
There are $N$ physical solutions to these equations given by
$\tilde{{\bf \Lambda}}=0$, $\tilde{{\bf p}}_{\nu_{\mu}}
\propto \hat{{\bf k}}$, and $\tilde{{\bf p}}_{\nu}=0$ for
$\nu\neq \nu_{\mu}$ with eigenvalue $\Omega^{2}_{\nu_{\mu}}
(k)$ for $\mu=1,\ldots N$.  There is also one unphysical
solution (it violates the gauge condition) given by
$\tilde{{\bf \Lambda}}\propto \hat{{\bf k}}$ and all of the
$\tilde{{\bf p}}_{\nu}$ being equal to zero.

We are now left with $2N+2$ transverse solutions.  We first
note that each transverse eigenvalue is two-fold degenerate.
This follows from the fact that if $M\overline{{\bf \lambda}}
=\omega^{2}\overline{{\bf \lambda}}$, then $M(K_{3N}
\overline{{\bf \lambda}})=\omega^{2}(K_{3N}
\overline{{\bf \lambda}})$, where
\begin{equation}
K_{3N}=\left(\begin{array}{ccc} K & 0 & \ldots \\
0 & K & \ldots \\ \vdots & \vdots &  \end{array}
\right),
\end{equation}
which can be verified by noting that if 
$\tilde{{\bf \Lambda}}$ and $\tilde{{\bf p}}_{\nu}$ satisfy
Eqs. (\ref{mode3d}), so do ${\bf k}\times \tilde{{\bf \Lambda}}$
and ${\bf k}\times \tilde{{\bf p}}_{\nu}$.  We choose the 
two eigenvectors $\overline{{\bf \lambda}}^{\mu 1}$ and
$\overline{{\bf \lambda}}^{\mu 2}$, which correspond to
the eigenvalue $\omega_{\mu}^{2}$, to be orthogonal 
in the sense that 
\begin{equation}
\langle \overline{{\bf \lambda}}^{\mu 1}|G
\overline{{\bf \lambda}}^{\mu 2}\rangle = 0.
\end{equation}
This, along with Eq. (\ref{orth1}) implies that 
\begin{equation}
\label{ortho}
\langle\overline{{\bf \lambda}}^{(\mu\sigma)}|G
\overline{{\bf \lambda}}^{(\mu^{\prime}\sigma^{\prime})}
\rangle
=\delta_{\mu\mu^{\prime}}\delta_{\sigma\sigma^{\prime}},
\end{equation}
which is our final orthonormality relation. Here,
due to isotropy,
the mode frequency $\omega_{\mu\sigma}$ does not depend
on the polarization index, $\sigma$, for the transverse 
modes.

We now express the fields in terms of the eigenvectors,
substitute them into the Hamiltonian density in Eq. (\ref{Hdensity}),
and integrate over the relevant n-dimensional volume.  In 
particular, we have that 
\begin{eqnarray}
{\bf \lambda}_{0}^{(\mu\sigma )}={\bf \Lambda}_{\mu\sigma} & 
\hspace{1cm} & {\bf \lambda}_{\nu}^{(\mu\sigma )}
= {\bf p}_{\mu\sigma }^{\nu} \\
{\bf \Pi}_{\mu\sigma}=-i\mu \omega_{\mu}
{\bf\Lambda}_{\mu\sigma} & \hspace{1cm} &
{\bf \pi}^{\nu}_{\mu\sigma}=\frac{-i\omega_{\mu}}
{\epsilon_{0}g_{\nu}}{\bf \lambda}_{\nu}^{(\mu\sigma)} .
\end{eqnarray}
Adding the requirement that 
$\overline{{\bf \lambda}}^{(\mu\sigma)}({\bf k})^{\ast}
=\overline{{\bf \lambda}}^{(\mu\sigma)}(-{\bf k})$ (it can be
shown that $M(-{\bf k})\overline{{\bf \lambda}}^
{(\mu\sigma)}({\bf k})^{\ast} = \omega^{2}_{\mu}(k)
\overline{{\bf \lambda}}^{(\mu\sigma)}({\bf k})^{\ast}$
which implies that $\overline{{\bf \lambda}}^{(\mu\sigma)}
({\bf k})^{\ast}$ is in the two-dimensional subspace spanned
by $\overline{{\bf \lambda}}^{(\mu\sigma)}(-{\bf k})$ for
$\sigma = 1,2$) and utilizing Eq. (\ref{ortho}), we find that
the terms of the form $\hat{a}_{\mu\sigma}({\bf k})
\hat{a}_{\mu^{\prime}\sigma^{\prime}}(-{\bf k})$ vanish giving
for the transverse modes
\begin{equation}
H_{\rm trans}=2(2\pi )^{n}\mu A\sum_{\mu =0}^{N}\sum_{\sigma =1}^{2}
\int d^{n}{\bf k} \langle \overline{{\bf \lambda}}^{(\mu\sigma)}
({\bf k})|G
\overline{{\bf \lambda}}^{(\mu\sigma)}({\bf k})\rangle 
\omega_{\mu}^{2}(k)\hat{a}_{\mu\sigma}^{\dagger}
({\bf k})\hat{a}_{\mu\sigma}({\bf k}) .
\end{equation}

In order to show that the Hamiltonian assumes the form given in
Eq. (\ref{H3D}) we need to prove that
\begin{equation}
\label{ccond}
2(2\pi )^{n}\mu A \langle\overline{{\bf \lambda}}^{(\mu\sigma)}
({\bf k})|G
\overline{{\bf \lambda}}^{(\mu\sigma)}({\bf k})\rangle 
\omega_{\mu}(k) = \hbar .
\end{equation}
Making use of Eqs. (\ref{Lambdamus}) and (\ref{pmus}) this condition becomes
\begin{equation}
\label{ccond2}
\left( 1+\sum_{\nu}\frac{k^{2}c^{2}g_{\nu}}
{(\Omega_{\nu}^{2}(k)
-\omega_{\mu}^{2})^{2}}\right)\frac{kv_{\mu\sigma}}
{\Omega_{\mu}} = 1 .
\end{equation}
This  agrees precisely with Eq. (\ref{vg2}) and is true even including phonon
dispersion ($\alpha_{\nu} \neq 0$).
However, as pointed out earlier, when there is phonon dispersion we cannot
interpret $v_{\mu\sigma}$ as the total group velocity - it only includes an
electromagnetic contribution, i.e., it is no longer equal to $\partial \omega
/\partial k$.

In order to complete the diagonalization of the Hamiltonian we must
consider the longitudinal modes.  The fields ${\bf \Lambda}$ and
${\bf \Pi}$ have no longitudinal components so that this part of
the diagonalization procedure involves only the fields 
${\bf p}_{\nu}$ and ${\bf \pi}_{\nu}$.  We find that
\begin{equation}
H_{\rm long}=(2\pi )^{3}\frac{2A}{\epsilon_{0}}\sum_{\mu =1}^{N}
\int d^{3}k \frac{\omega_{\mu 0}^{2}(k)}{g_{\mu}}
{\bf p}_{\mu 0}^{\mu}({\bf k})^{\ast}\cdot
{\bf p}_{\mu 0}^{\mu}({\bf k})\hat{a}_{\mu 0}^{\dagger}
\hat{a}_{\mu 0}({\bf k}) ,
\end{equation}
where we have made use of the fact that ${\bf p}_{\mu 0}^{\nu}
({\bf k}) = \delta_{\mu \nu}{\bf p}_{\mu 0}^{\mu}({\bf k})$,
and we note that $\omega_{\mu 0}(k)=\Omega_{\mu}(k)$.  The 
Hamiltonian assumes the expected form,
\begin{equation}
H_{\rm long}=\sum_{\mu =1}^{N}
\int d^{3}k \hbar\omega_{\mu 0}(k)
\hat{a}_{\mu 0}^{\dagger}
\hat{a}_{\mu 0}({\bf k}) ,
\end{equation}
when explicit expressions
for the vectors, ${\bf p}_{\mu 0}^{\mu}({\bf k})$ from 
Eq. (\ref{pmulong}) are used.

The final Hamiltonian, which is the sum of $H_{\rm trans}$ 
and $H_{\rm long}$, has ($3N+2$) mode operators
for each value of ${\bf k}$, and can be written in the form%
\begin{equation}
H=\sum\limits_{\mu=0}^{N} \sum\limits_{\sigma=0'}^{2}
\int \hbar \omega _{\mu\sigma}({k})\widehat{a}_{\mu\sigma}^{\dagger
}({\bf k})%
\widehat{a}_{\mu\sigma}({\bf k})\,d^n{\bf k}\ . 
\end{equation}
Here  the lower limit $\sigma=0'$ excludes the
combination of $\mu=0$ and $\sigma=0$, which would imply a longitudinal
polariton. Also, there is a requirement of having ($N+1$)
distinct roots for this form to be valid. The corresponding field operators
(in the full three-dimensional case) have the expansions:%
\begin{eqnarray}
\widehat{\bf D}&=
&i\sum\limits_{\mu,\sigma=1,2}\int d^3{\bf k}\,
\left[ \frac{\hbar {kv^{\scriptscriptstyle EM}_{\mu}(k)\varepsilon(\omega _{\mu}(k))}}{4\pi
 }\right] ^{1/2}
{\bf e}_{\sigma}({\bf k})\widehat{a}_{\mu\sigma}({\bf k})
e^{i{\bf k\cdot x}} + hc \nonumber \\
\widehat{\bf E}^{\perp}&=
&i\sum\limits_{\mu,\sigma=1,2}\int d^3{\bf k}\,
\left[ \frac{\hbar {kv^{\scriptscriptstyle EM}_{\mu}(k)}}{4\pi
\varepsilon(\omega _{\mu}(k))}\right] ^{1/2} 
{\bf e}_{\sigma}({\bf k})\widehat{a}_{\mu\sigma}({\bf k})
e^{i{\bf k\cdot x}} + hc \nonumber \\
\widehat{\bf B}&=&-i\sum\limits_{\mu,\sigma=1,2}\int d^3{\bf k}\,
\left[ \frac{\hbar\mu k{v^{\scriptscriptstyle EM}_{\mu}(k)}}{4\pi
 }\right] ^{1/2}{\bf u}_{\sigma}({\bf k})
 \widehat{a}_{\mu\sigma}({\bf k})
 e^{i{\bf k\cdot x}} + hc\ . 
\end{eqnarray}

Here we have introduced the electric field mode 
${\bf e}({\bf k}) = {\bf k}\times {\bf u}({\bf k})/|{\bf k}|$ to simplify the expansion.
We note, as in the
one-dimensional case, that the transverse field expansion for the electric field is
simply derived from the displacement field by using the frequency-dependent permittivity.
The main feature introduced by the dispersion is the replacement of a frequency term $\omega$,
that would normally appear in the expansion coefficients, by a new term with
the same units, but equal to $k{v^{\scriptscriptstyle EM}_{\mu}(k)}$ instead. We do not
give the expansion for the longitudinal part of the electric field here explicitly,
except to point out that it is equal to $-{\bf P}^{\parallel}/\varepsilon_0$.

\section{Summary}

A simple theory of a one-dimensional, dispersive waveguide was introduced,
including a polarizable model of the medium with $N$ discrete localized
resonances. This can be thought of as a limiting case of an ideal insulator,
in which the polarization field is due to localized electrons at each atomic
location. The theory is exactly equivalent to the usual classical theory of
a dispersive dielectric medium, in the sense that it results in the
Sellmeir equations for the refractive index. These are well-known to lead
to an excellent fit to the classical dispersion properties of transparent
media, and have the usual causality requirements automatically satisfied.
The theory was quantized and a set of $N+1$ mode operators introduced, for
the polaritons in
each branch of the dispersion relation,
provided there were $N+1$ distinct, positive roots.
 In this case, the mode-expansion
has a universal and simple form, only depending on the group velocity.

This  model is necessarily causal, and implements the
causality requirements through band-gaps, rather than isolated poles. It does
omit many important correction factors that occur in practice. In particular,
our mode expansion neglects scattering off inhomogeneities.
For this reason, transmission inside the transmission band is essentially
lossless.
It also omits
 nonlinearities
due to phonon-phonon, photon-photon, and photon-phonon interactions,
which are responsible for additional non-electromagnetic damping
of the polaritons.
 However,
these effects can certainly be added to the Hamiltonian once a mode
expansion is established.

Next,
a quantum theory of an isotropic 
n-dimensional dispersive waveguide was introduced,
with $n=1,2$ or $3$.
Without any additional phonon/exciton dispersion,
the theory is exactly equivalent to the  classical Drude-Lorentz theory of
a dispersive dielectric medium.
The 
complete n-dimensional
theory was quantized, and a set of ($3N+2$) mode operators introduced, for
each branch of the dispersion relation;
again, with the restriction of distinct, positive roots. 
As in the one-dimensional case, the mode expansion
 depends on the permittivity and the
electromagnetic group velocity, in the case of transverse polaritons. However,
the group velocity factor in this case  is modified
to include only the electromagnetic
component of the group velocity.

\section*{Acknowledgment}
This work was supported in part by the 
Australian Research Council, and by the US
National Science Foundation under
grants INT-9602515 and PHY94-07194.

\end{document}